\documentclass[twocolumn]{aastex7}
\linenumbers

\usepackage{float}
\usepackage{comment}
\usepackage{amsmath, amssymb, amsthm}
\usepackage{booktabs}
\usepackage{graphicx}
\usepackage{dcolumn} 
\usepackage{lineno}

\newcommand{\water}{H$_\textrm{2}$O}
\newcommand{\methane}{CH$_{4}$}
\newcommand{\co}{CO}
\newcommand{\coo}{CO$_{2}$}
\newcommand{\ammonia}{NH$_{3}$}
\newcommand{\hhs}{H$_{2}$S}
\newcommand{\teff}{$T_\textrm{eff}$}
\newcommand{\logg}{$\log_{10}(g)$}

\newcommand{\mum}{$\mu$m}

\newcommand{\fwater}{$f_\textrm{H$_{2}$O}$}
\newcommand{\fmethane}{$f_\textrm{CH$_{4}$}$}
\newcommand{\fco}{$f_\textrm{CO}$}
\newcommand{\fcoo}{$f_\textrm{CO$_{2}$}$}

\newcommand{\fhhs}{$f_\textrm{H$_{2}$S}$}

\begin{document}

\title{A Comprehensive Atmospheric Retrieval Analysis of 22 James Webb Space Telescope Spectral Energy Distributions of Cool Brown Dwarfs}

\author[0009-0009-4489-0192]{Harshil Kothari} \email{harshil177@icloud.com}
\affiliation{Ritter Astrophysical Research Center, Department of Physics \& Astronomy, University of Toledo, 2801 W. Bancroft St., Toledo, OH 43606, USA}

\author[0000-0003-4269-260X]{Michael C. Cushing} \email{x}
\affiliation{Ritter Astrophysical Research Center, Department of Physics \& Astronomy, University of Toledo, 2801 W. Bancroft St., Toledo, OH 43606, USA}

\author[0000-0002-6721-1844]{Samuel A. Beiler} \email{x}
\affiliation{Ritter Astrophysical Research Center, Department of Physics \& Astronomy, University of Toledo, 2801 W. Bancroft St., Toledo, OH 43606, USA}

\author[0000-0001-6627-6067]{Channon Visscher} \email{x}
\affiliation{Chemistry \& Planetary Sciences, Dordt University, Sioux Center,
51250, IA, USA.}

\author[0000-0002-5251-2943]{Mark S. Marley} \email{x}
\affiliation{Lunar and Planetary Laboratory, University of Arizona, 1629 E. University Boulevard, Tucson, AZ 85721, USA}

\author[0000-0003-4600-5627]{Ben Burningham} \email{x}
\affiliation{Centre for Astrophysics Research, School of Physics, Astronomy and Mathematics, University of Hertfordshire, Hatfield AL10 9AB}

\author[0000-0002-4249-864X]{Adam C. Schneider} \email{x}
\affiliation{United States Naval Observatory, Flagstaff Station, 10391 West Naval Observatory Road, Flagstaff, AZ 86005, USA}

\author[0000-0003-4269-260X]{J. Davy Kirkpatrick} \email{x}
\affiliation{IPAC, Mail Code 100-22, Caltech, 1200 E. California Boulevard, Pasadena, CA 91125, USA}

\begin{abstract}
We present a uniform atmospheric retrieval analysis of 22 late-T and Y-type brown dwarfs within 20 pc, observed with the James Webb Space Telescope NIRSpec PRISM and MIRI LRS. This dataset provides the first continuous $\sim$0.95–12 $\mu$m spectroscopic coverage of late-T and Y-type brown dwarfs, which in turn enables precise constraints on their thermal structures and volume mixing ratios (VMRs) of H$_{2}$O, CH$_{4}$, CO, CO$_{2}$, NH$_{3}$, H$_{2}$S, K, Na, and PH$_{3}$. We find positive correlations between the VMR of \water\ and \methane, and \co\ and \coo, consistent with thermochemical equilibrium chemistry. Using the VMRs, we derive atmospheric metallicity, which is positively correlated with \water\ and \methane, showing \water\ and \methane\ trace oxygen and carbon content, respectively, allowing us to effectively measure (O/H)$_\textrm{bulk}$ and (C/H)$_\textrm{bulk}$. We also report tentative PH$_3$ detections in roughly half the sample, suggesting potential vertical mixing or non-equilibrium chemistry. Apart from chemical properties, we retrieve masses and radii spanning $\sim$6–77 $\mathcal{M}^\textrm{N}_\textrm{Jup}$ and $\sim$0.66–1.53 $\mathcal{R}^\textrm{N}_\textrm{Jup}$, respectively. We compare the derived $\log_{10}(g)$ ($\sim$ 4–5.5 [cm s$^{-2}$]) and \teff\ ($\sim$350–1100 K) with Sonora Bobcat evolutionary models and find an age range of 0.4 to 10 Gyr amongst the sample. Comparing our retrieved thermal profiles with the Elf-Owl forward model thermal profiles, we find a systematic difference between the two, likely arising due to the difference in chemistry treatment.
\end{abstract}


\section{Introduction} \label{sec:introduction}

Brown dwarfs, particularly the ones at the coldest effective temperatures ($<$600 K), present a fundamental challenge for atmospheric characterization. Their spectral energy distributions (SEDs) peak at $\sim$5 $\mu$m, and yet for years spectroscopic observations were largely constrained to the near-infrared (1--2.5 $\mu$m) \citep[e.g.,][]{Schneider_2015,2016ApJ...824....2L} due to the difficulty of observing from the ground at $\lambda > 2.5 \mu$m and the lack of space-based spectroscopic capabilities before the launch of the James Webb Space Telescope (JWST).

\citet{Miles_2020} presented spectra of seven brown dwarfs from the Gemini-GNIRS, revealing CO features in the 3–5 \mum\ wavelength range. Beyond these data, mid-infrared (2.5--30 $\mu$m) spectroscopy of very late-T and Y dwarfs remains sparse. Only a small number ($\sim$20) of objects later than T6 have spectral coverage from the Spitzer Space Telescope, primarily using the IRS instrument over $\sim$5--15 $\mu$m, and those observations are limited in wavelength range and signal-to-noise. As a result, detailed spectroscopic constraints across the full near- to mid-infrared spectral energy distribution remain rare for the coldest brown dwarfs.

The James Webb Space Telescope (JWST) has transformed the cold brown dwarf landscape. With a continuous spectral coverage from $\sim$1–28~\mum, JWST captures all of the major carbon-, nitrogen-, and oxygen-bearing molecules (\water, \methane, \co, \coo, \ammonia) that make up the atmospheres of the coldest brown dwarfs. The unprecedented wavelength coverage from the JWST enables us to precisely determine the chemical abundances, isotopic ratios, carbon-to-oxygen ratios, metallicity, and thermal structures of these cold atmospheres \citep{Barrado_2023, Kothari_2024, Lew_2024}. For the first time, it is possible to draw robust conclusions about the chemical diversity and physical processes governing these cool atmospheres, paving the way for a deeper understanding of their formation and evolution.

There are two techniques to model the atmospheres of brown dwarfs.  The first, called forward modeling,  solves for the thermal structure (i.e. the run of temperature with pressure) of the atmosphere for a give set of fundamental parameters (e.g. metallicity, surface gravity, effective temperature, etc.) assuming things such as thermochemical equilibrium, a chemistry scheme, and one-dimensional radiative-convective equilibrium.  The resulting model emergent spectrum can then be compared to observations in order to infer the atmospheric properties of the brown dwarfs.  The second technique, known as atmospheric retrieval, was originally developed to study planetary atmospheres in the solar system \citep[e.g.,][]{Chahine_1968} and aims to infer the thermal structure
of the atmosphere along with the abundances of the major absorbers directly from the data. 

Retrievals have been successfully applied to L- and T-type brown dwarf spectra, yielding new insights into their chemistry, cloud formation, and elemental abundances \citep{Line_2014, Line_2017, Burningham_2017}. Previous studies relied primarily on near-infrared spectra obtained with facilities such as the Hubble Space Telescope and ground-based instruments, which probe only a limited fraction of the emergent flux from these $\lesssim$500 K atmospheres. At such low temperatures, key molecular absorbers (e.g., NH$_3$, CO$_2$, and other disequilibrium species) exhibit prominent features at longer wavelengths, leading to strong degeneracies in thermal structure and chemical abundances when constrained by near-infrared data alone \citep{Zalesky_2019, Zalesky_2022}. Consequently, retrievals of late-T and Y dwarfs based solely on near-IR spectra have carried substantial uncertainties, limiting robust conclusions about their atmospheric properties.

Recent JWST studies of individual Y dwarfs have already  uncovered evidence for strong vertical mixing \citep{Kothari_2024} and disequilibrium chemistry in the coldest atmospheres \citep{Beiler_2024, Faherty_2024}, and in some cases suggest temperature structures that deviate from forward-model predictions \citep{Kothari_2024}. While these JWST spectral analyses have provided unprecedented insights, they have largely focused on single objects, leaving open questions about population-level trends across the late-T and Y-dwarf regime.

In this paper, we present a comprehensive and uniform atmospheric retrieval analysis of 22 late-T and Y-type brown dwarfs, spanning spectral types from T6 to Y1. Using low-resolution spectra from JWST, we investigate the thermal profiles, chemical abundances, and physical properties of these objects. This dataset represents the largest sample of cold brown dwarfs analyzed with JWST to date, providing a unique opportunity to explore population-level trends and variation in their atmospheric properties, allowing us to understand the diversity of brown dwarf atmospheres. In \S \ref{sec:data}, we describe the observations and data reduction process. In \S \ref{sec:method}, we detail the retrieval framework employed to analyze the spectra, including the model assumptions and parameter space explored. The retrieved results are presented and discussed in \S \ref{sec:results}, where we highlight key trends and correlations across the sample. Finally, in \S \ref{sec:discussion}, we discuss our findings with respect to thermochemical equilibrium predictions and forward models, and discuss the implications for understanding the atmospheric physics and chemistry of cold brown dwarfs.

\section{The Spectra} \label{sec:data}

We have analyzed spectra of 22 brown dwarfs within 20 pc of the Sun with spectral types ranging from T6 to Y1. The spectra of these brown dwarfs were observed as a part of JWST General Observers (GO) program \#2302. For more details about the sample selection please refer to  \cite{Beiler_2024_b}.  

The spectra were obtained using the Near Infrared Spectrograph \citep[hereafter NIRSpec,][]{Jakobsen_2022}, which covers 0.6--5.3~$\mu$m with a resolving power ($R=\lambda/\Delta \lambda$) of 30--300, and the Mid-Infrared Instrument \citep[hereafter MIRI,][]{Rieke_2015}, which covers~5--14 $\mu$m with a resolving power of 50--200. The NIRSpec and MIRI data were reduced as described in \citet{Beiler_2024_b}. Briefly, the data were processed using the JWST Science Calibration Pipeline (Version 1.12.5) with Calibration Reference Data System (CRDS) version 11.17.6 and 1193.pmap CRDS context to assign the reference files. The standard pipeline stages were executed to produce background-subtracted, wavelength- and flux-calibrated two-dimensional products and extracted one-dimensional spectra.

\citet{Beiler_2024_b} used Spitzer/IRAC Channel 2 ([4.5]) photometry from \citet{Kirkpatrick_2012} and MIRI F1000W ($\lambda_\textrm{pivot}=9.954~\mu$m) photometry, obtained as part of the same GO program, to calibrate the absolute flux density levels of the NIRSpec and MIRI spectra, respectively, to an overall precision of $\sim3\%$. \citeauthor{Beiler_2023} then created a continuous spectrum of each object spanning $\sim0.95$--12~$\mu$m by merging the NIRSpec and MIRI spectra between 5 and 5.3~$\mu$m, where the wavelength coverage overlaps. We adopt the reduced spectra from \citet{Beiler_2024_b} for this analysis.

\section{Retrieval Methodology} \label{sec:method}

Atmospheric retrieval iteratively generates and compares model spectra to the observed spectrum using Bayesian inference, with the goal of estimating the joint posterior probability density function (PDF) of a set of model parameters $(\boldsymbol{\Theta})$ given an observed spectrum $(\boldsymbol{f}_\lambda)$. For our retrieval analysis, we use the Nested Sampling version of the Brewster retrieval framework \citep{Burningham_2017}, which is set up as described in \citet{Kothari_2024}. 

The vector $\boldsymbol{\Theta}$ includes parameters for: the mixing ratio of the gases, the thermal profile of the atmosphere, physical characteristics like mass, radius, and systematic parameters (e.g., error inflation and wavelength offset). (see Table \ref{priors}. for more details). The retrieval framework uses a sampling technique (e.g. MCMC, Nested Sampling, etc.) that iteratively optimizes the model to fit the data by computing posterior PDF for the model parameters ($\boldsymbol{\Theta}$) using Bayes' theorem:

\begin{equation}
p(\boldsymbol{\Theta}|\boldsymbol{f}_\lambda) = \frac{p(\boldsymbol{\Theta}) \mathcal{L}(\boldsymbol{f}_\lambda|\boldsymbol{\Theta})}{p(\boldsymbol{f}_\lambda)},
\end{equation}

\noindent where:
\begin{itemize}
    \item \(p(\boldsymbol{\Theta}|\boldsymbol{f}_\lambda)\) is the posterior PDF that represents the updated belief about the model parameters after incorporating the observed spectrum $(\boldsymbol{f}_\lambda)$.
    \item \(p(\boldsymbol{\Theta})\) is the prior PDF, encoding prior knowledge or assumptions about the parameters.
    \item \(\mathcal{L}(\boldsymbol{f}_\lambda|\boldsymbol{\Theta})\) is the likelihood function, which quantifies the probability of observing the spectrum $(\boldsymbol{f}_\lambda)$ given a set of model parameters \(\boldsymbol{\Theta}\).
    \item \(p(\boldsymbol{f}_\lambda)\) is the evidence, which allows us to compare different retrieval models.
\end{itemize}

\subsection{Generative Model} \label{sec:Generative Model}

The model-predicted flux densities are calculated at each wavelength, \(\mathcal{M}_\lambda(\lambda_i)\) as:

\begin{equation}
\mathcal{M}_\lambda(\lambda_i) = \left(\frac{R}{d}\right)^2 \left[\boldsymbol{I}(\lambda_i) * \boldsymbol{\mathcal{F}}_\lambda(\boldsymbol{\theta}_\textrm{atm}, \lambda_j)\right],
\end{equation}

\noindent where:
\begin{itemize}
    \item \(\boldsymbol{\mathcal{F}}_\lambda(\boldsymbol{\theta}_\textrm{atm}, \lambda_j)\) are the emergent model flux densities at the top of the atmosphere given a set of atmospheric parameters (\(\boldsymbol{\theta}_\textrm{atm}\) = $\log_{10} (f_i)$, $M$, $R$, and \(T_\textrm{Knot-i}\)), calculated by using a two-stream source function technique described in \citet{Toon_1989}. \(\lambda_j\) is equal to \(\lambda_k\) + \(\Delta \lambda\), where \(\lambda_k\) is the wavelength at which the model emergent flux density is calculated and  \(\Delta \lambda\) is a parameter that accounts for uncertainty in wavelength calibration.
    \item \(\boldsymbol{I}(\lambda_i)\) is the instrument profile, modeled as Gaussian, which accounts for the variable resolving power of the observed spectrum at each wavelength (\(\lambda_i\)) \citep{Beiler_2023, Kothari_2024}.
    \item \(R/d\) is a scaling factor to scale the model spectrum as it is observed from Earth. 
\end{itemize}

Each datum in the observed spectrum is modeled probabilistically as:

\begin{equation}
F_\lambda(\lambda_i) = \mathcal{M}_\lambda(\lambda_i) + \epsilon(\lambda_i),
\end{equation}

\noindent where $F_\lambda(\lambda_i)$ is a random variable denoting the flux density at wavelength $\lambda_i$ and \(\epsilon(\lambda_i)\) is a Gaussian random parameter with zero mean and variance given by \(\sigma^2(\lambda_i)\). Therefore, $\boldsymbol \Theta = \{ \boldsymbol \theta_\textrm{atm}, d, b, \Delta \lambda\}$.

\subsection{Likelihood Function}

Assuming the noise in an observed spectrum is Gaussian and the spectral points are independent, the likelihood function of the data given the model is given by:

\begin{equation}
\ln \mathcal{L}(\boldsymbol{f}_\lambda|\boldsymbol{\Theta}) = -\frac{1}{2} \sum_{i=1}^n \left\{ \frac{[f_{\lambda, i} - \mathcal{M}_\lambda (\lambda_i)]^2}{\sigma^2(\lambda_i)} + \ln[2\pi \sigma^2(\lambda_i)] \right\},
\end{equation}

\noindent where \(f_{\lambda, i}\) is the observed flux density at wavelength \(\lambda_i\), \(\mathcal{M}_\lambda(\lambda_i)\) is the predicted model flux density at wavelength \(\lambda_i\), described in \S\ref{sec:Generative Model}, and \(\sigma^2(\lambda_i)\) is the variance at \(\lambda_i\). This variance is modeled as:

\begin{equation}
\sigma^2(\lambda_i) = s^2(\lambda_i) + 10^b,
\label{variance_eq}
\end{equation}

\noindent where \(s(\lambda_i)\) is the standard error of the measured flux density returned by JWST pipeline at $\lambda_i$ , and \(b\) is a tolerance parameter to account for unmodeled systematic uncertainties \citep[e.g.,][]{Hogg_2010, Foreman_2013, Burningham_2017, Kothari_2024}.

\subsection{Atmospheric Model}

For the atmospheric model, we divide the atmosphere into 64 layers \footnote{We performed retrievals with 100 atmospheric layers for three objects (regular, $Y$-peculiar, and $YJH$-peculiar; see \S\ref{modelspectra}) and found no statistically significant impact on the retrieval results; all inferred parameters are consistent within 1$\sigma$ between the two model configurations.} with pressure levels ranging from \(10^{-4}\) to \(10^{2.3}\) bar in steps of 0.1 dex. The thermal profile of the atmosphere is parameterized using a 5-knot \footnote{We  performed a retrieval of a regular object (see \S\ref{modelspectra}) with a 9-knot thermal profile parameterization and found no statistically significant impact on the retrieval results; all inferred parameters remain consistent within 1$\sigma$.} interpolating spline, with knots {located at $\log_{10}$P = $-$4 [top], $-$2.425, $-$0.85 [middle], 0.725, and 2.3 [bottom], respectively.

The atmosphere is assumed to be cloud-free \footnote{We adopt a cloud-free atmospheric model for two reasons: 1) Our goal is to identify the minimum number of parameters required to adequately reproduce the observed spectra. Introducing additional free parameters, such as a gray cloud deck, increases model flexibility and also introduces strong degeneracies with the temperature structure and molecular abundances, particularly at the S/N ratio and spectral resolution of the current data. 2) We employ a "free" thermal profile model which avoids or minimizes cloud impact on the overall retrieved results. A dedicated exploration of cloud parameters in the retrieval model is therefore deferred to future work, where the statistical significance of clouds and their impact on retrieved abundances and degeneracies with the thermal structure can be explored in a systematic way.}, with the following gases contributing to the opacity: H\(_2\), He, H\(_2\)O, CH\(_4\), CO, CO\(_2\), NH\(_3\), H\(_2\)S, K, Na, and PH\(_3\). These species are chosen because they are predicted to dominate the gas-phase opacity budget \citep[e.g.,][]{Lodders_2006}.  In particular, H\(_2\) and He constitute the bulk of the gas by number while H\(_2\)O, CH\(_4\), and NH\(_3\) are the primary infrared absorbers in this regime.   CO, CO\(_2\), H\(_2\)S, and PH\(_3\) act as important trace species that can probe disequilibrium chemistry and Na and K provide strong pressure-broadened alkali resonance wings in the red optical and near infrared. The continuum opacity is contributed by H\(_2\) and He via collision-induced absorption (H\(_2\)-H\(_2\), H\(_2\)-He). The non-continuum cross-sections are computed from linelists that are sampled at an equivalent resolving power of $R \sim 10,000$ \footnote{We performed a retrieval of a regular object (see \S\ref{modelspectra}) with linelists computed at R=30,000 and found no statistically significant impact on the retrieved results; all inferred parameters remain consistent within 1$\sigma$.}. The linelists are adopted from the following sources: \citet{Tennyson_2018} for \water, \citet{Hargreaves_2020} for \methane, \citet{Li_2015} for CO,  \citet{Huang_2014} for \coo, \citet{Yurchenko_2011} for \ammonia, \citet{Azzam_2015} for \hhs, \citet{Freedman_2008, Freedman_2014A} for Na and K, and \citet{Marley_2021} for PH\(_3\). Pressure broadening coefficients for H$_2$$–$He atmospheres are computed using the methodology described in \citet{Gharib-Nezhad_2021A}. The line width is calculated from the Van der Waals broadening theory for collisions with H\(_2\) molecules using the coefficient tabulated in the VALD3 database when available, or from the full theory otherwise. For the H\(_2\)–H\(_2\) and H\(_2\)–He CIA, we adopt tabulated absorption coefficients from the same opacity database used in our previous retrieval work (see \citealt{Burningham_2017} for details of the implementation). For a comprehensive discussion of the opacity treatment and cross-section computation, see \citet{Burningham_2017}. The volume mixing ratios of the gases $(f_i)$ are treated as free parameters, which are assumed to be constant throughout the atmosphere (i.e. uniform-with-pressure), along with physical properties like mass (\(M\)) and radius (\(R\)), which are used to determine the surface gravity.

\begin{deluxetable*}{c c c}[t!] 
\centering
\tablecaption{Parameter Priors \label{priors}}
\tablehead{
\colhead{Parameter} &
\colhead{Description} &
\colhead{Prior\tablenotemark{a}}}
\startdata
    $\log_{10}(f_i)$\tablenotemark{b,c} & Gas Volume Mixing Ratio  & $\mathcal{U}(-12,0)$
    \\
    $M$ & Mass [$\mathcal{M}^\textrm{N}_\textrm{Jup}$] & $\mathcal{U}(0.1, 80)$
    \\
    $R$ & Radius [$\mathcal{R}^\mathrm{N}_{e\mathrm{J}}$] & $\mathcal{U}(0.5, 2)$ 
    \\
    $\Delta \lambda$ & Wavelength shift [$\mu$m] & $\mathcal{U}(-0.01, 0.01)$ 
    \\
    $10^b$ & Error inflation &  $\mathcal{U}({0.01}\times \textrm{min}(\sigma_{i}^{2})), {100} \times \textrm{max}(\sigma_{i}^{2}))$ 
    \\
    $d$ & Distance [pc]\tablenotemark{d} & $\mathcal{N}$($\mu$, $\sigma^{2})$
    \\
    $T_{\textrm{Knot-i}}$\tablenotemark{e} & Thermal profile knots [K] & $\mathcal{U}(0, 5000)$ 
\enddata

\tablenotetext{a}{$\mathcal{U}(\alpha, \beta)$ denotes a uniform distribution between $\alpha$ and $\beta$ while $\mathcal{N}(\mu, \sigma^2)$ denotes a normal distribution with a mean of $\mu$ and a variance of $\sigma^2$.}

\tablenotetext{b}{Our retrieval model retrieves volume mixing ratio (the number density of the species divided by the total number number density of the gas) for H$_{2}$O, CH$_{4}$, CO, CO$_{2}$, NH$_{3}$, H$_{2}$S, K, Na, and PH$_{3}$.}

\tablenotetext{c}{All volume mixing ratios are reported as the log of the ratio, and the remainder of the gas is assumed to be H$_{2}$-He (1$-$$\sum_i f_i$). Assuming a solar abundance of 91.2\% of number of atoms of H and 8.7\% of number of atoms of He \citep{Asplund_2009}, 84\% of the volume mixing ratio is from H$_{2}$ and 16\% is from He for the remainder gas.}

\tablenotetext{d}{The priors for distance were adopted from \citet{Beiler_2024_b}.}

\tablenotetext{e}{This profile does not allow for temperature inversions, i.e., each $T_{\textrm{Knot}}$ must have a decreasing value with decreasing pressure.}
\end{deluxetable*}

\section{Results} \label{sec:results}

\begin{figure*}[htb!]
    \centering
    \includegraphics[width=\linewidth]{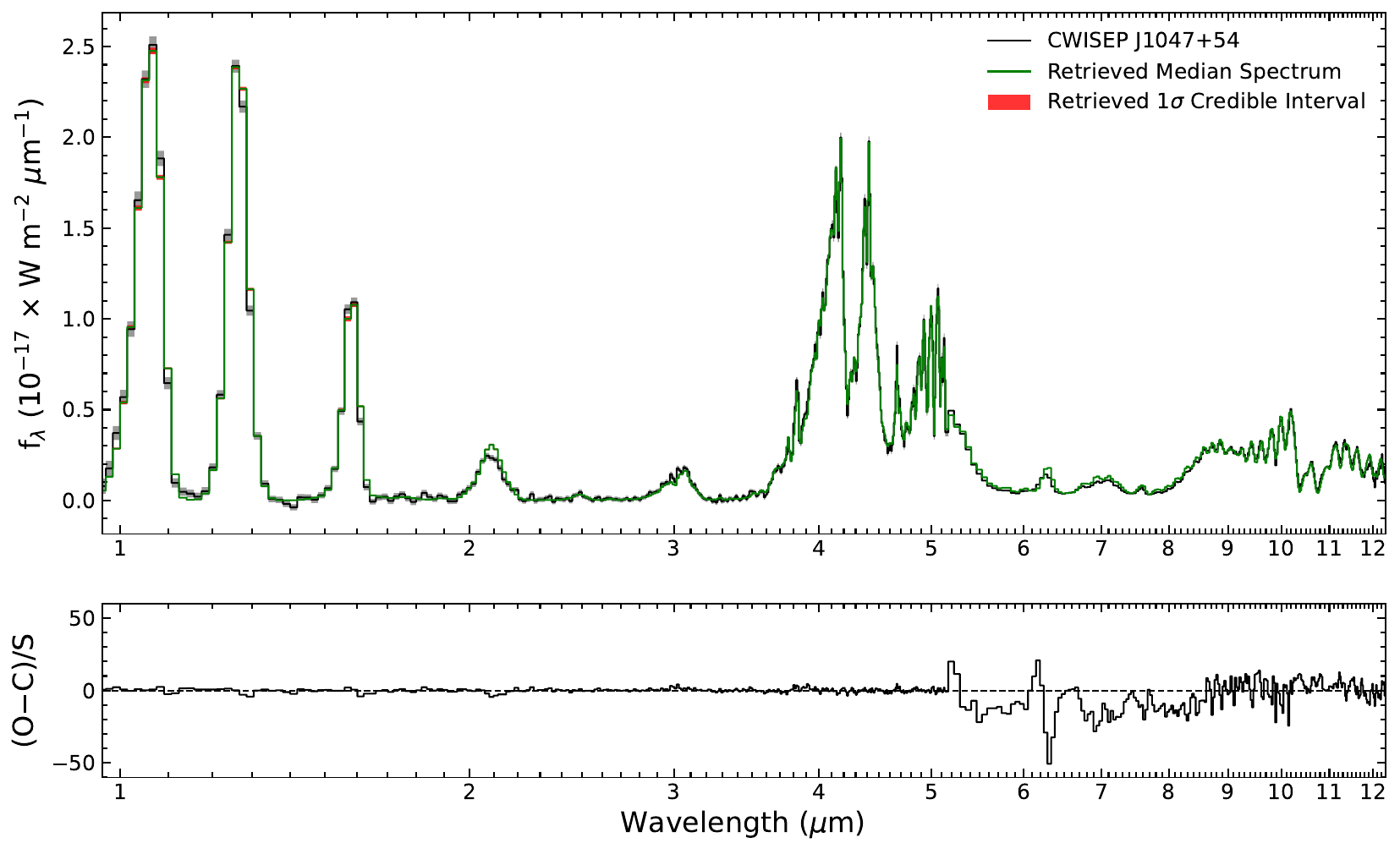}
    \caption{The top panel shows the observed JWST spectrum (NIRSpec [PRISM] + MIRI low-resolution spectrometer [LRS]) of CWISEP J1047+54 (spectral type Y1) in black covering $\sim$0.96–12.2 $\mu$m with 1$\sigma$ uncertainties in grey in units of f$_{\lambda}$. The retrieved median spectrum is shown in green and the red region shows the 1$\sigma$ central credible interval around the median spectrum. The bottom panel shows the difference between the observed (O) and retrieved median spectrum (C) divided by the observed spectrum uncertainties (S).}
    \label{1047_spectrum}
\end{figure*}

\begin{figure*}[htb!]
    \centering
    \includegraphics[width=\linewidth]{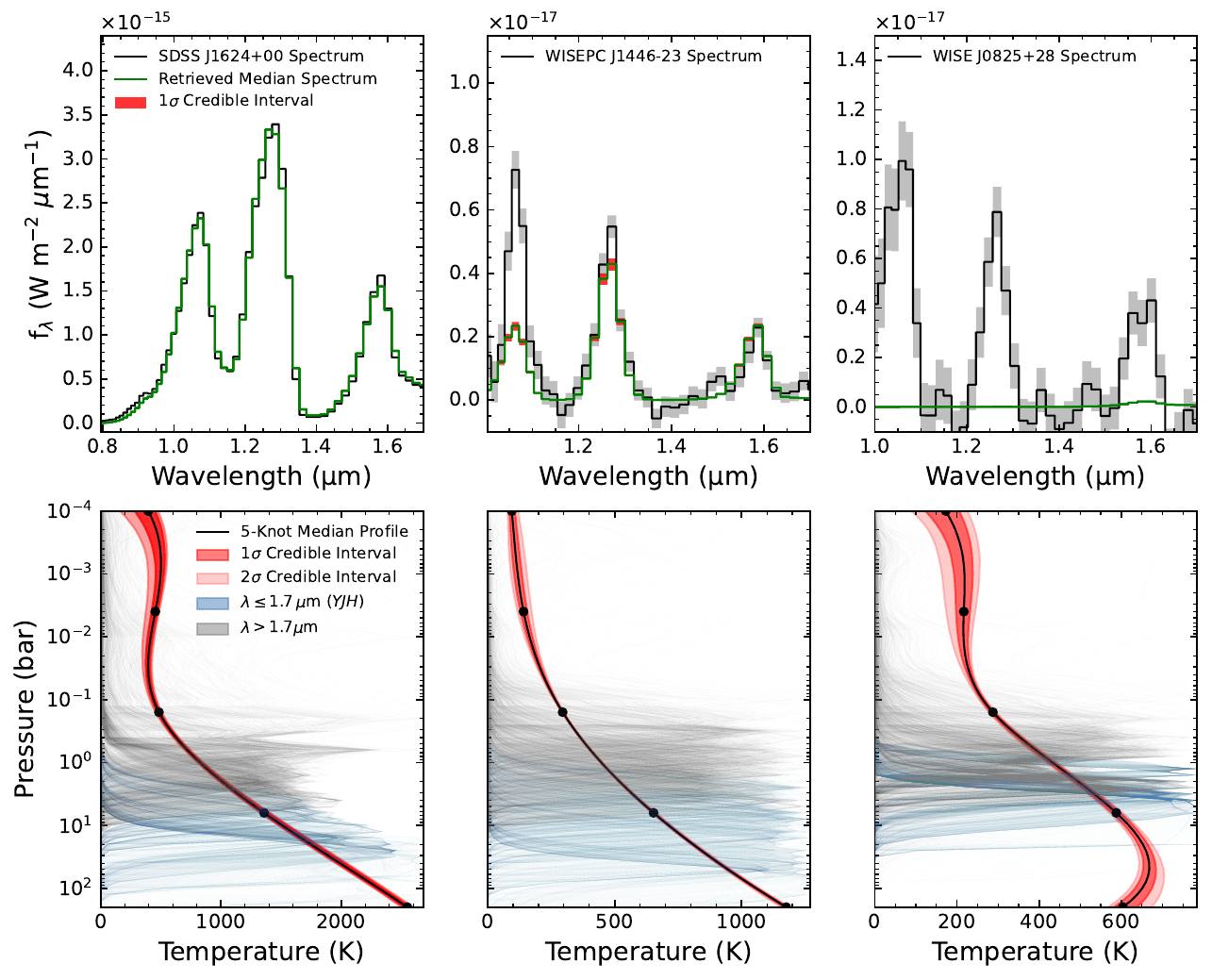}
    \caption{Top Panels: Observed versus retrieved spectra for three categories—regular, $Y$-peculiar, and $YJH$-peculiar—are shown. The black line indicates the observed spectrum, with grey shading denoting the uncertainties. The green line shows the median retrieved spectrum, and the red shading represents the 1$\sigma$ central credible interval from the retrieval. Bottom Panels: Retrieved thermal profiles corresponding to each model-fit category. The black line shows the median thermal profile, with dark and light red regions indicating the 1$\sigma$ and 2$\sigma$ central credible intervals, respectively. The grey and blue lines represent normalized contribution functions for different wavelength regions (blue: $<$1.7\mum\ and grey: $>$1.7\mum), and the black dots represent the five retrieved temperature-pressure knots.}
    \label{spec_tp}
\end{figure*}

The end result of the atmospheric retrieval is a 19-dimensional PDF, which can be marginalized to obtain 1D posterior PDF for each parameter. In the following sections, we discuss both the chemical properties ($\log_{10} (f_i)$; see Table \ref{chem_prop}), physical properties (mass, radius, distance, surface gravity, and effective temperature; see Table \ref{phys_prop}), and chemical ratios ((C/O)$_\mathrm{gas}$, (C/O)$_\mathrm{bulk}$, [M/H]$_\textrm{gas}$, [M/H]$_\textrm{bulk}$, (O/H)$_\mathrm{gas}$, (O/H)$_\mathrm{bulk}$, (C/H)$_\mathrm{gas}$, and (S/H)$_\mathrm{gas}$; see Table \ref{chem_ratio}) based on the 1D marginalized posterior PDFs for all 22 objects in our sample.

\subsection{Retrieved Model Spectra}
\label{modelspectra}

Figure \ref{1047_spectrum} shows a typical example of our retrieved model fits to the data for CWISEP J104756.81+545741.6 (hereafter we abbreviate target names as HHMM$\pm$DD, e.g. CWISEP J1047+54 [SpT: Y1]). The model spectra, generated using the retrieved posterior parametric values, generally fit the observed spectra well, especially for the NIRSpec PRISM data ($\sim$0.95 to 5.2 \mum). Any large discrepancies in the model fit to the data are confined beyond 5 \mum, which is the MIRI LRS spectrum. There are two possible reasons for these discrepancies: 1) Incorrect wavelength-dependent resolving power [$R$($\lambda$)] of the MIRI LRS instrument which can over-smooth or under-smooth key spectral features, and 2) The MIRI LRS spectra have higher S/N ratio than the NIRSpec PRISM spectra, resulting in smaller observational uncertainties. As a consequence, even small mismatches between the model and the data become more apparent in the residuals. These larger normalized residuals therefore reflect the higher precision of the MIRI data rather than a substantially poorer overall fit of the model.\footnote{We performed an additional test in which the retrieval model included two independent tolerance parameters (eq~\ref{variance_eq}), one for the NIRSpec/PRISM and another for the MIRI/LRS part of the spectrum. We conducted this test for two regular objects as well as peculiar objects, and found the residuals remained the same, i.e., the residuals increased where the PRISM and LRS spectra are stitched just like they do with only one tolerance parameter, with no statistically significant changes in the credible intervals.}

Model spectra of certain objects fit the data poorly at near-infrared wavelengths ($\sim$0.95 to 1.7~$\mu$m). Therefore, we classify the objects from our sample into three categories based on the quality of the model spectrum fit to the data at near-infrared wavelengths. The top three panels in Figure \ref{spec_tp} illustrates the model fits to the observed spectra for representative objects from each classification category: regular (15 objects), $Y$-peculiar (WISEPC J1405+55, CWISEP J1446--23, and WISEA J2354+02), and $YJH$-peculiar (WISE J0535--75, WISE J0825+28, WISEPA J1541--22, and WISE J2209+27). 

For a regular object (e.g. Figure \ref{spec_tp} [top left]), the model fits the data in the near-infrared well. However, the model spectra for three $Y$-peculiar objects do not fit the observed spectra well in the $Y$ band (e.g. Figure \ref{spec_tp} [top middle]). For the four $YJH$-peculiar objects, the model spectra predict little to no flux in the near-infrared part of the spectrum (e.g. Figure \ref{spec_tp} [top right]). We surmise the spectral mismatches arise primarily to the relatively low S/N in the wavelength regions spanning the $Y$, $J$, and $H$ bands. As a result, the high S/N MIRI LRS data is dominating the likelihood evaluations which preferentially optimizes the parameters to fit the MIRI portion of the spectrum. As a result, for objects with relatively low S/N in $Y$, $J$, and $H$ band compared to the MIRI LRS spectrum, the retrieval is effectively driven by the longer wavelengths till an adequate fit to the MIRI data is achieved, given our Nested Sampling convergence criteria. 

The absence of visibly broad credible intervals (red) at wavelengths shorter than 1.7\mum\ for the $YJH$-peculiar objects reflects the tight posterior constraints on the retrieved parameters that define the model spectra. These narrow intervals therefore do not necessarily indicate a superior fit in this region, but rather arise from the dominance of the higher-precision mid-infrared data in shaping the posterior distributions.

Despite the discrepancies in the near-infrared bands, the retrieved model spectra reproduces the respective observed spectrum (beyond $\sim$1.7~$\mu$m) with equally good fits across all three categories. In other words, while the $Y$-peculiar and $YJH$-peculiar objects show mismatches in the $Y$, $J$, and $H$ bands due to low S/N, their model-fits remain as robust as those of the regular objects at longer wavelengths.

\subsection{The Thermal Profile}
\label{sec:5knotprofile}

Figure \ref{spec_tp} [Bottom] shows the retrieved thermal profiles for each model spectral fit category, with the full set of retrieved thermal profiles for all the 22 objects provided in Figures  \ref{tp1}–\ref{tp4}. In all these figures, the solid black lines represent the median retrieved profiles, computed using the median values of the parameters, while the red shaded regions around the median profile indicate the 1$\sigma$ (16th–84th percentiles) and 2$\sigma$ (2.4th–97.6th percentiles) central credible intervals \footnote{A Bayesian central credible interval gives the range of values in a parameter's posterior distribution that contain $\alpha$\% of the probability.  In contrast, a frequentist $\alpha$\% confidence interval means that $\alpha$\% of a large number of confidence intervals computed in the same way would contain the true value of the parameter.}, respectively. The black dots represent the pressure level of the knots where the temperature is being retrieved. The grey and blue lines are contribution functions, which represents the relative contribution of different layers of the atmosphere to the observed radiation, i.e., they indicate the pressure levels the spectrum probes for each object at different wavelengths. The blue contribution functions represent the flux from the $Y,J,$ and $H$ bands ($\sim$1$-$1.7~\mum), and the grey contribution functions represent the flux from the rest of the wavelength coverage ($>$1.7~\mum).

Most of the retrieved thermal profiles exhibit a general trend of decreasing temperature with decreasing pressure, with a few profiles exhibiting a temperature reversal at the top of the atmosphere (typically less than 10$^{-3}$ bars) for seven (WISEPA J0313+78 [T8.5], WISE J2102$-$44 [T9], WISE J0359$-$54 [Y0], WISE J0734$-$71 [Y0], WISE J1206+84 [Y0], WISEPC J2056+14 [Y0], WISEPC J1405+55 [Y0.5]). These temperature reversals are due to a slight increase in the contribution functions at less than 10$^{-3}$ bars primarily from the longer wavelength absorption features like the $\sim$7 to 8 \mum\ \methane\ band and $\sim$10 to 11 \mum\ \ammonia\ band. \citet{Faherty_2024} reported a connection between temperature inversions and auroral activity in brown dwarfs. However, when we increased the number of spline knots used to parameterize the thermal profile, the retrieved thermal structure no longer exhibited a thermal reversal at the top of the atmosphere. This demonstrates that the apparent thermal reversal obtained with the 5-knot spline parameterization is not a physically robust feature, but rather a relic of the 5-knot spline interpolation. \footnote{We do not adopt a larger number of spline knots for the thermal profile parameterization for the entire sample because doing so would substantially increases the computational cost of the retrievals while yielding no statistically significant improvement in the inferred atmospheric properties.}

The retrieved thermal profiles of the four $YJH$-peculiar objects (WISE J0825+28 (see Figure \ref{spec_tp} [Bottom third]), WISE J0535--75, WISEPA J1541--22, and WISE J2209+27; see Figure \ref{tp3} \& \ref{tp4}) are irregular and wiggly compared to the retrieved thermal profiles for the regular and the $Y$-peculiar objects. The $YJH$-peculiar objects thermal profiles generally exhibit an increase in temperature deep in the atmosphere (up to 100 bar), followed by a monotonic decrease in temperature up to $\sim$0.1 bar. At that point, the profiles undergo a temperature reversal up to $\sim$0.001 bar, before decreasing again at lower pressures. The temperature reversal at high-pressures results in a lack of near-infrared flux in the model spectrum, particularly in the $Y$, $J$, and $H$ bands, which probe deeper parts of the atmosphere, as evidenced by the top and bottom right panels in Figure \ref{spec_tp}. This is further supported by the absence of contribution functions from the deepest part of our model atmosphere (around 10$^2$ bars), which are probed by the wavelength range covered by the $Y$, $J$, and $H$ bands. Therefore, the curved shape of thermal profiles deep in the atmosphere for $YJH$-peculiar objects is non-physical and is a result of poor S/N in the $Y$, $J$, and $H$ bands of the observed spectra.

Among the three $Y$-peculiar objects, the retrieved thermal profiles of two (WISEPC J1405+55 and WISEA J2354+02) deviate noticeably from the regular trend, particularly at pressures below $\sim$0.1 bar. The profile for WISEPC J1405+55 (see Figure \ref{spec_tp} [bottom second]) begins to converge toward an isothermal shape at pressures less than 0.01 bar, with wider central credible intervals than those of regular objects. The profile for WISEA J2354+02 (see Figure \ref{tp4}) shows a slight temperature reversal between 0.01 and 0.001 bar, followed by a monotonic decrease at lower pressures, along with broader central credible intervals than other $Y$-peculiar and regular profiles. These isothermal-like and curvy features at low pressures are artifacts of the spline interpolation and should not be interpreted as physical, as supported by the contribution functions indicating the spectra are most sensitive to $\sim$0.1–100 bar.

\subsection{Chemical Trends}
\label{mixingratios}

\begin{deluxetable*}{lcccccccccccc}
\tablecaption{Retrieved Chemical Properties \label{chem_prop}}
\tablehead{
\colhead{Object Name} &
\colhead{SpT} &
\colhead{Peculiarity} &
\colhead{$\log_{10}(f_\textrm{H$_{2}$O})$} &
\colhead{$\log_{10}(f_\textrm{CH$_{4}$})$} &
\colhead{$\log_{10}(f_\textrm{CO})$} &
\colhead{$\log_{10}(f_\textrm{CO$_{2}$})$} &
\colhead{$\log_{10}(f_\textrm{NH$_{3}$})$} &
\colhead{$\log_{10}(f_\textrm{H$_{2}$S})$} &
\colhead{$\log_{10}(f_\textrm{Na})$} &
\colhead{$\log_{10}(f_\textrm{K})$} &
\colhead{$\log_{10}(f_\textrm{PH$_{3}$})$} 
}
\startdata
SDSSJ1624+00& T6& Regular& $-3.31^{+0.02}_{-0.02}$
 & $-3.57^{+0.03}_{-0.03}$& $-4.51^{+0.04}_{-0.04}$
 & $-7.16^{+0.05}_{-0.05}$& $-8.83^{+1.86}_{-1.80}$
 & $-8.95^{+1.82}_{-1.87}$& $-5.37^{+0.03}_{-0.03}$
 & $-6.71^{+0.02}_{-0.02}$& $-9.90^{+1.29}_{-1.37}$  \\ 
WISEJ1501-40& T6& Regular& $-3.03^{+0.01}_{-0.01}$
 & $-3.24^{+0.01}_{-0.01}$& $-3.92^{+0.02}_{-0.02}$
 & $-6.38^{+0.02}_{-0.02}$& $-5.48^{+0.09}_{-0.07}$
 & $-9.08^{+1.94}_{-2.03}$& $-5.25^{+0.02}_{-0.02}$
 & $-6.43^{+0.02}_{-0.02}$& $-9.53^{+1.65}_{-1.59}$  \\ 
SDSSpJ1346-00& T6.5& Regular& $-3.22^{+0.02}_{-0.02}$
 & $-3.48^{+0.02}_{-0.04}$& $-3.71^{+0.03}_{-0.04}$
 & $-6.37^{+0.04}_{-0.04}$& $-5.25^{+0.08}_{-0.07}$
 & $-8.98^{+1.90}_{-1.91}$& $-5.52^{+0.09}_{-0.07}$
 & $-6.42^{+0.03}_{-0.03}$& $-7.63^{+2.35}_{-0.47}$  \\ 
ULASJ1029+09& T8& Regular& $-2.88^{+0.01}_{-0.01}$
 & $-2.91^{+0.01}_{-0.01}$& $-3.99^{+0.01}_{-0.01}$
 & $-6.45^{+0.01}_{-0.01}$& $-4.87^{+0.03}_{-0.03}$
 & $-8.83^{+1.71}_{-1.81}$& $-5.33^{+0.02}_{-0.04}$
 & $-6.94^{+0.04}_{-0.03}$& $-6.79^{+0.06}_{-0.05}$  \\ 
WISEJ0247+37& T8& Regular& $-2.92^{+0.01}_{-0.01}$
 & $-3.00^{+0.01}_{-0.01}$& $-4.92^{+0.01}_{-0.01}$
 & $-7.34^{+0.01}_{-0.01}$& $-4.72^{+0.02}_{-0.01}$
 & $-8.26^{+1.34}_{-1.19}$& $-7.51^{+1.02}_{-0.72}$
 & $-6.45^{+0.01}_{-0.01}$& $-7.82^{+0.77}_{-0.24}$  \\ 
WISEJ0430+46& T8& Regular& $-3.37^{+0.01}_{-0.02}$
 & $-3.76^{+0.02}_{-0.03}$& $-6.00^{+0.02}_{-0.02}$
 & $-8.60^{+0.05}_{-0.06}$& $-4.84^{+0.02}_{-0.03}$
 & $-4.62^{+0.08}_{-0.07}$& $-9.82^{+1.33}_{-1.32}$
 & $-7.20^{+0.02}_{-0.03}$& $-9.82^{+1.32}_{-1.04}$  \\ 
WISEPAJ1959-33& T8& Regular& $-2.98^{+0.02}_{-0.03}$
 & $-3.06^{+0.03}_{-0.04}$& $-3.96^{+0.02}_{-0.02}$
 & $-6.51^{+0.04}_{-0.04}$& $-4.77^{+0.04}_{-0.05}$
 & $-9.03^{+1.96}_{-1.97}$& $-5.51^{+0.04}_{-0.04}$
 & $-6.89^{+0.04}_{-0.03}$& $-7.01^{+0.10}_{-0.09}$  \\ 
WISEPAJ0313+78& T8.5& Regular
 & $-3.05^{+0.03}_{-0.03}$& $-3.09^{+0.04}_{-0.04}$
 & $-5.02^{+0.02}_{-0.02}$& $-7.52^{+0.05}_{-0.04}$
 & $-4.65^{+0.04}_{-0.04}$& $-8.63^{+2.20}_{-2.32}$
 & $-9.54^{+1.61}_{-1.67}$& $-6.85^{+0.03}_{-0.03}$
 & $-7.29^{+0.08}_{-0.07}$  \\ 
WISEA2159-48& T9& Regular& $-3.11^{+0.02}_{-0.02}$
 & $-3.40^{+0.03}_{-0.04}$& $-5.69^{+0.02}_{-0.03}$
 & $-8.33^{+0.06}_{-0.06}$& $-4.74^{+0.04}_{-0.04}$
 & $-5.08^{+4.20}_{-0.36}$& $-9.64^{+1.61}_{-1.76}$
 & $-6.92^{+0.02}_{-0.03}$& $-7.59^{+0.09}_{-0.08}$  \\ 
WISEJ2102-44& T9& Regular& $-2.87^{+0.01}_{-0.02}$
 & $-2.88^{+0.01}_{-0.02}$& $-4.71^{+0.01}_{-0.02}$
 & $-7.15^{+0.02}_{-0.02}$& $-4.46^{+0.02}_{-0.02}$
 & $-8.81^{+1.77}_{-1.90}$& $-9.57^{+1.32}_{-1.56}$
 & $-6.73^{+0.01}_{-0.01}$& $-7.06^{+0.05}_{-0.05}$  \\ 
WISEJ0359-54& Y0& Regular& $-3.08^{+0.02}_{-0.03}$
 & $-3.40^{+0.03}_{-0.04}$& $-5.19^{+0.02}_{-0.03}$
 & $-8.04^{+0.04}_{-0.05}$& $-4.61^{+0.04}_{-0.04}$
 & $-4.09^{+0.05}_{-0.06}$& $-9.68^{+1.50}_{-1.49}$
 & $-10.26^{+1.15}_{-1.16}$& $-8.65^{+1.57}_{-0.37}$  \\ 
WISEJ0734-71& Y0& Regular& $-3.00^{+0.02}_{-0.02}$
 & $-3.19^{+0.02}_{-0.03}$& $-5.05^{+0.02}_{-0.02}$
 & $-7.72^{+0.03}_{-0.03}$& $-4.71^{+0.03}_{-0.03}$
 & $-3.89^{+0.04}_{-0.04}$& $-9.00^{+1.98}_{-1.82}$
 & $-7.50^{+0.07}_{-0.06}$& $-7.79^{+0.11}_{-0.09}$  \\ 
WISEJ1206+84& Y0& Regular& $-2.85^{+0.02}_{-0.02}$
 & $-3.06^{+0.03}_{-0.03}$& $-4.25^{+0.02}_{-0.02}$
 & $-6.98^{+0.03}_{-0.03}$& $-4.62^{+0.03}_{-0.03}$
 & $-3.69^{+0.04}_{-0.04}$& $-9.40^{+1.73}_{-1.87}$
 & $-7.58^{+0.10}_{-0.08}$& $-7.52^{+0.09}_{-0.08}$  \\ 
WISEJ2209+27& Y0& Regular& $-3.02^{+0.02}_{-0.02}$
 & $-3.14^{+0.03}_{-0.03}$& $-5.21^{+0.02}_{-0.02}$
 & $-7.90^{+0.03}_{-0.03}$& $-4.54^{+0.03}_{-0.03}$
 & $-4.77^{+1.19}_{-0.19}$& $-2.68^{+0.09}_{-0.08}$
 & $-7.35^{+2.80}_{-2.86}$& $-8.40^{+0.33}_{-0.19}$  \\ 
WISEPCJ2056+14& Y0& $YJH$-Pec.
 & $-3.00^{+0.03}_{-0.03}$& $-3.27^{+0.04}_{-0.04}$
 & $-4.67^{+0.03}_{-0.03}$& $-7.49^{+0.04}_{-0.04}$
 & $-4.84^{+0.05}_{-0.05}$& $-3.97^{+0.06}_{-0.06}$
 & $-9.50^{+1.66}_{-1.66}$& $-7.29^{+0.05}_{-0.04}$
 & $-7.67^{+0.09}_{-0.08}$  \\ 
WISEJ0825+28& Y0.5& $Y$-Pec.& $-3.04^{+0.02}_{-0.02}$
 & $-3.26^{+0.03}_{-0.03}$& $-5.47^{+0.03}_{-0.03}$
 & $-8.13^{+0.04}_{-0.05}$& $-4.74^{+0.03}_{-0.04}$
 & $-8.76^{+2.05}_{-2.15}$& $-2.36^{+0.06}_{-0.06}$
 & $-7.30^{+2.98}_{-3.07}$& $-10.08^{+1.18}_{-1.09}$  \\ 
WISEPCJ1405+55& Y0.5& Regular
 & $-3.01^{+0.02}_{-0.02}$& $-3.29^{+0.02}_{-0.03}$
 & $-5.68^{+0.02}_{-0.02}$& $-8.59^{+0.04}_{-0.05}$
 & $-4.47^{+0.03}_{-0.03}$& $-3.98^{+0.03}_{-0.04}$
 & $-9.85^{+1.27}_{-1.44}$& $-10.53^{+0.87}_{-0.99}$
 & $-8.10^{+0.08}_{-0.08}$  \\ 
CWISEPJ1047+54& Y1& $Y$-Pec.& $-2.51^{+0.01}_{-0.01}$
 & $-2.60^{+0.01}_{-0.01}$& $-4.22^{+0.01}_{-0.01}$
 & $-6.57^{+0.01}_{-0.01}$& $-4.00^{+0.01}_{-0.01}$
 & $-4.13^{+0.11}_{-0.06}$& $-8.94^{+0.80}_{-0.73}$
 & $-9.11^{+0.44}_{-0.36}$& $-8.79^{+0.90}_{-0.96}$  \\ 
CWISEPJ1446-23& Y1& $Y$-Pec.& $-2.54^{+0.01}_{-0.01}$
 & $-2.64^{+0.01}_{-0.01}$& $-4.66^{+0.01}_{-0.01}$
 & $-7.08^{+0.02}_{-0.01}$& $-4.21^{+0.02}_{-0.01}$
 & $-3.32^{+0.02}_{-0.02}$& $-8.14^{+0.69}_{-0.63}$
 & $-8.89^{+0.58}_{-0.50}$& $-9.73^{+0.95}_{-1.06}$  \\ 
WISEAJ2354+02& Y1& $YJH$-Pec.
 & $-2.89^{+0.02}_{-0.02}$& $-3.15^{+0.02}_{-0.02}$
 & $-5.52^{+0.02}_{-0.02}$& $-8.00^{+0.03}_{-0.04}$
 & $-4.41^{+0.02}_{-0.02}$& $-4.03^{+0.04}_{-0.05}$
 & $-9.61^{+1.46}_{-1.52}$& $-10.01^{+1.14}_{-1.13}$
 & $-8.27^{+0.26}_{-0.15}$  \\ 
WISEJ0535-75& Y1& $YJH$-Pec.& $-3.22^{+0.03}_{-0.03}$
 & $-3.50^{+0.03}_{-0.04}$& $-6.29^{+0.04}_{-0.04}$
 & $-9.03^{+0.10}_{-0.10}$& $-4.80^{+0.04}_{-0.04}$
 & $-7.95^{+2.42}_{-2.28}$& $-2.89^{+0.09}_{-0.07}$
 & $-7.74^{+2.68}_{-2.75}$& $-9.42^{+1.42}_{-0.71}$  \\ 
WISEPAJ1541-22& Y1& $YJH$-Pec.
 & $-3.07^{+0.02}_{-0.03}$& $-3.22^{+0.03}_{-0.03}$
 & $-5.37^{+0.03}_{-0.03}$& $-7.95^{+0.04}_{-0.04}$
 & $-4.81^{+0.04}_{-0.04}$& $-8.32^{+2.30}_{-2.34}$
 & $-2.28^{+0.06}_{-0.06}$& $-7.22^{+2.96}_{-3.02}$
 & $-9.65^{+1.40}_{-1.04}$  
\enddata
\end{deluxetable*}

\begin{deluxetable*}{l c c c c c c r r}[htb!]
\tablecolumns{7}
\centering
\tablecaption{Physical Properties \label{phys_prop}}
\tablehead{
\colhead{Object Name} & 
\colhead{SpT} &
\colhead{Peculiarity} &
\colhead{Mass \tablenotemark{a}} & 
\colhead{Radius \tablenotemark{a}} &
\colhead{Distance \tablenotemark{a}} & 
\colhead{$\log_{10}(g)$ \tablenotemark{b}} &
\colhead{T$_\textrm{eff}$ \tablenotemark{b}} &
\colhead{T$_\textrm{eff}^\textrm{uni}$ \tablenotemark{c}} \\
\colhead{} &
\colhead{} &
\colhead{} &
\colhead{($\mathcal{M}^\textrm{N}_\textrm{Jup}$) \tablenotemark{d}} &
\colhead{($\mathcal{R}^\textrm{N}_\textrm{Jup}$) \tablenotemark{e}} &
\colhead{(pc)} & 
\colhead{[cm s$^{-2}$]} &
\colhead{(K)} &
\colhead{(K)}
}
\startdata
SDSSJ1624+00& T6& Regular& $17^{+3}_{-2}$
 & $0.73^{+0.01}_{-0.01}$& $10.90^{+0.11}_{-0.10}$
 &$4.89^{+0.08}_{-0.05}$&$1037^{+7}_{-7}$
 &$987^{+22}_{-60}$  \\ 
WISEJ1501-40& T6& Regular& $29^{+2}_{-2}$
 & $0.66^{+0.02}_{-0.02}$& $13.77^{+0.34}_{-0.35}$
 &$5.21^{+0.02}_{-0.02}$&$1010^{+16}_{-16}$
 &$897^{+27}_{-48}$  \\ 
SDSSpJ1346-00& T6.5& Regular& $19^{+2}_{-4}$
 & $0.96^{+0.02}_{-0.02}$& $14.51^{+0.36}_{-0.34}$
 &$4.71^{+0.05}_{-0.08}$&$978^{+15}_{-15}$
 &$1044^{+26}_{-55}$  \\ 
ULASJ1029+09& T8& Regular& $67^{+3}_{-4}$
 & $0.82^{+0.01}_{-0.01}$& $14.55^{+0.13}_{-0.13}$
 &$5.39^{+0.02}_{-0.02}$&$766^{+4}_{-5}$
 &$784^{+31}_{-50}$  \\ 
WISEJ0247+37& T8& Regular& $63^{+2}_{-2}$
 & $0.83^{+0.01}_{-0.01}$& $15.38^{+0.23}_{-0.17}$
 &$5.36^{+0.01}_{-0.01}$&$679^{+6}_{-6}$
 &$669^{+22}_{-35}$  \\ 
WISEJ0430+46& T8& Regular& $19^{+1}_{-2}$
 & $0.78^{+0.02}_{-0.01}$& $10.46^{+0.21}_{-0.19}$
 &$4.88^{+0.02}_{-0.05}$&$560^{+7}_{-7}$
 &$525^{+16}_{-26}$  \\ 
WISEPAJ1959-33& T8& Regular& $49^{+8}_{-11}$
 & $0.94^{+0.02}_{-0.02}$& $11.95^{+0.23}_{-0.22}$
 &$5.15^{+0.07}_{-0.08}$&$762^{+9}_{-10}$
 &$756^{+23}_{-39}$  \\ 
WISEPAJ0313+78& T8.5& Regular& $32^{+6}_{-7}$
 & $0.88^{+0.02}_{-0.02}$& $7.38^{+0.13}_{-0.13}$
 &$5.01^{+0.09}_{-0.08}$&$591^{+6}_{-7}$
 &$805^{+24}_{-43}$  \\ 
WISEA2159-48& T9& Regular& $26^{+3}_{-6}$
 & $0.87^{+0.03}_{-0.03}$& $13.66^{+0.44}_{-0.46}$
 &$4.94^{+0.06}_{-0.08}$&$571^{+11}_{-12}$
 &$598^{+17}_{-27}$  \\ 
WISEJ2102-44& T9& Regular& $56^{+4}_{-4}$
 & $0.85^{+0.01}_{-0.01}$& $10.74^{+0.10}_{-0.11}$
 &$5.28^{+0.03}_{-0.03}$&$590^{+4}_{-4}$
 &$585^{+17}_{-27}$  \\ 
WISEJ0359-54& Y0& Regular& $12^{+1}_{-2}$
 & $0.96^{+0.02}_{-0.02}$& $13.60^{+0.29}_{-0.29}$
 &$4.49^{+0.04}_{-0.09}$&$454^{+6}_{-6}$
 &$565^{+20}_{-26}$  \\ 
WISEJ0734-71& Y0& Regular& $13^{+1}_{-2}$
 & $0.93^{+0.02}_{-0.02}$& $13.43^{+0.24}_{-0.25}$
 &$4.56^{+0.04}_{-0.05}$&$481^{+5}_{-6}$
 &$468^{+13}_{-23}$  \\ 
WISEJ1206+84& Y0& Regular& $18^{+2}_{-3}$
 & $1.07^{+0.02}_{-0.02}$& $11.83^{+0.25}_{-0.23}$
 &$4.60^{+0.06}_{-0.06}$&$456^{+6}_{-6}$
 &$493^{+14}_{-24}$  \\ 
WISEJ2209+27& Y0& Regular& $9^{+1}_{-1}$
 & $1.06^{+0.01}_{-0.01}$& $6.18^{+0.07}_{-0.06}$
 &$4.28^{+0.05}_{-0.05}$&$350^{+3}_{-3}$
 &$499^{+14}_{-24}$  \\ 
WISEPCJ2056+14& Y0& $YJH$-Pec.& $6^{+1}_{-1}$
 & $1.08^{+0.01}_{-0.02}$& $7.10^{+0.09}_{-0.09}$
 &$4.09^{+0.09}_{-0.09}$&$463^{+4}_{-4}$
 &$510^{+14}_{-25}$  \\ 
WISEJ0825+28& Y0.5& $Y$-Pec.& $11^{+2}_{-3}$
 & $1.11^{+0.02}_{-0.02}$& $6.55^{+0.07}_{-0.07}$
 &$4.33^{+0.07}_{-0.09}$&$363^{+3}_{-3}$
 &$371^{+9}_{-13}$  \\ 
WISEPCJ1405+55& Y0.5& Regular& $8^{+1}_{-1}$
 & $0.93^{+0.01}_{-0.01}$& $6.33^{+0.07}_{-0.06}$
 &$4.38^{+0.04}_{-0.05}$&$405^{+3}_{-3}$
 &$406^{+10}_{-16}$  \\ 
CWISEPJ1047+54& Y1& $Y$-Pec.& $41^{+1}_{-1}$
 & $0.88^{+0.01}_{-0.01}$& $14.04^{+0.15}_{-0.17}$
 &$5.12^{+0.02}_{-0.01}$&$415^{+3}_{-3}$
 &$412^{+10}_{-17}$  \\ 
CWISEPJ1446-23& Y1& $Y$-Pec.& $18^{+1}_{-1}$
 & $0.82^{+0.02}_{-0.02}$& $9.44^{+0.16}_{-0.13}$
 &$4.81^{+0.03}_{-0.02}$&$375^{+4}_{-5}$
 &$415^{+20}_{-22}$  \\ 
WISEAJ2354+02& Y1& $YJH$-Pec.& $14^{+1}_{-2}$
 & $0.88^{+0.02}_{-0.01}$& $7.70^{+0.12}_{-0.10}$
 &$4.65^{+0.05}_{-0.05}$&$371^{+4}_{-4}$
 &$366^{+14}_{-14}$  \\ 
WISEJ0535-75& Y1& $YJH$-Pec.& $22^{+3}_{-4}$
 & $1.53^{+0.04}_{-0.04}$& $14.61^{+0.33}_{-0.32}$
 &$4.37^{+0.06}_{-0.07}$&$395^{+6}_{-6}$
 &$432^{+11}_{-18}$  \\ 
WISEPAJ1541-22& Y1& $YJH$-Pec.& $12^{+2}_{-2}$
 & $1.14^{+0.02}_{-0.02}$& $5.99^{+0.06}_{-0.06}$
 &$4.37^{+0.08}_{-0.08}$&$381^{+3}_{-3}$
 &$526^{+16}_{-26}$  
\enddata

\tablenotetext{a}{Mass, Radius, and Distance are retrieved parameters which are used to calculate surface gravity ($g$) and T$_\textrm{eff}$.}

\tablenotetext{b}{$\log_{10}$$(g)$ \& T$_\textrm{eff}$ are derived parameters calculated using equation \ref{gravity_eq} and \ref{teff_eq}, respectively.}

\tablenotetext{c}{Reported effective temperatures from \citet{Beiler_2024_b}, calculated using a uniform 1$-$10 Gyr age distribution assumption as preferred by \citet{Kirkpatrick_2024}.}

\tablenotetext{d}{$ \mathcal{M}^\textrm{N}_\textrm{Jup}$ is Jupiter’s nominal mass of  1.898 $\times$ 10$^{27}$ kg [assuming $G = 6.67430 \times 10^{-11} \textrm{ m}^3 \textrm{ kg}^{-1} \textrm{ s}^{-2}$] \citep{Mamajek_2015}.}

\tablenotetext{e}{$\mathcal{R}^\mathrm{N}_{e\mathrm{J}}$ is Jupiter’s nominal equatorial radius of 7.1492 $\times$ 10$^{7}$ m \citep{Mamajek_2015}.} 

\end{deluxetable*}

\begin{deluxetable*}{lcccccccccccr}
\centering
\tablecaption{Derived Chemical Ratios\label{chem_ratio}}
\tablehead{
\colhead{Object Name} &
\colhead{SpT} &
\colhead{Peculiarity} &
\colhead{(C/O)$_\textrm{gas}$} &
\colhead{(C/O)$_\textrm{bulk}$\tablenotemark{a}} &
\colhead{[M/H]$_\textrm{gas}$} &
\colhead{[M/H]$_\textrm{bulk}$\tablenotemark{a}} &
\colhead{(O/H)$_\textrm{gas}$} &
\colhead{(O/H)$_\textrm{bulk}$\tablenotemark{a}} &
\colhead{(C/H)$_\textrm{gas}$} &
\colhead{(S/H)$_\textrm{gas}$} 
}
\startdata
SDSSJ1624+00& T6& Regular&$0.540^{+0.002}_{-0.002}$
 &$0.450^{+0.001}_{-0.001}$&$-0.24^{+0.02}_{-0.02}$
 &$-0.18^{+0.03}_{-0.02}$&$-3.51^{+0.02}_{-0.02}$
 &$-3.43^{+0.02}_{-0.02}$&$-3.75^{+0.03}_{-0.02}$
 &$-9.17^{+1.82}_{-1.87}$  \\ 
WISEJ1501-40& T6& Regular&$0.535^{+0.001}_{-0.001}$
 &$0.446^{+0.001}_{-0.001}$&$0.09^{+0.01}_{-0.01}$
 &$0.15^{+0.01}_{-0.01}$&$-3.20^{+0.01}_{-0.01}$
 &$-3.11^{+0.01}_{-0.01}$&$-3.39^{+0.01}_{-0.01}$
 &$-9.31^{+1.94}_{-2.03}$  \\ 
SDSSpJ1346-00& T6.5& Regular
 &$0.540^{+0.003}_{-0.002}$&$0.450^{+0.002}_{-0.001}$
 &$-0.02^{+0.02}_{-0.02}$&$0.04^{+0.02}_{-0.02}$
 &$-3.32^{+0.02}_{-0.02}$&$-3.23^{+0.02}_{-0.02}$
 &$-3.50^{+0.02}_{-0.03}$&$-9.21^{+1.90}_{-1.91}$  \\ 
ULASJ1029+09& T8& Regular&$0.506^{+0.001}_{-0.001}$
 &$0.426^{+0.001}_{-0.001}$&$0.29^{+0.01}_{-0.01}$
 &$0.36^{+0.01}_{-0.01}$&$-3.07^{+0.01}_{-0.01}$
 &$-2.94^{+0.01}_{-0.01}$&$-3.10^{+0.01}_{-0.01}$
 &$-9.06^{+1.71}_{-1.81}$  \\ 
WISEJ0247+37& T8& Regular&$0.513^{+0.001}_{-0.001}$
 &$0.431^{+0.001}_{-0.001}$&$0.20^{+0.00}_{-0.00}$
 &$0.27^{+0.00}_{-0.00}$&$-3.14^{+0.01}_{-0.01}$
 &$-3.03^{+0.00}_{-0.01}$&$-3.22^{+0.01}_{-0.01}$
 &$-8.49^{+1.34}_{-1.19}$  \\ 
WISEJ0430+46& T8& Regular&$0.558^{+0.001}_{-0.001}$
 &$0.463^{+0.001}_{-0.001}$&$-0.35^{+0.01}_{-0.02}$
 &$-0.30^{+0.01}_{-0.02}$&$-3.59^{+0.01}_{-0.02}$
 &$-3.53^{+0.01}_{-0.02}$&$-3.99^{+0.02}_{-0.03}$
 &$-4.85^{+0.08}_{-0.07}$  \\ 
WISEPAJ1959-33& T8& Regular&$0.514^{+0.002}_{-0.002}$
 &$0.432^{+0.002}_{-0.002}$&$0.18^{+0.03}_{-0.03}$
 &$0.25^{+0.03}_{-0.03}$&$-3.16^{+0.02}_{-0.03}$
 &$-3.04^{+0.02}_{-0.03}$&$-3.23^{+0.03}_{-0.04}$
 &$-9.25^{+1.96}_{-1.97}$  \\ 
WISEPAJ0313+78& T8.5& Regular
 &$0.505^{+0.002}_{-0.002}$&$0.426^{+0.002}_{-0.002}$
 &$0.09^{+0.03}_{-0.03}$&$0.16^{+0.04}_{-0.03}$
 &$-3.28^{+0.03}_{-0.03}$&$-3.15^{+0.03}_{-0.03}$
 &$-3.31^{+0.04}_{-0.04}$&$-8.85^{+2.20}_{-2.32}$  \\ 
WISEA2159-48& T9& Regular&$0.546^{+0.002}_{-0.002}$
 &$0.454^{+0.002}_{-0.001}$&$-0.07^{+0.02}_{-0.03}$
 &$-0.02^{+0.02}_{-0.03}$&$-3.33^{+0.02}_{-0.02}$
 &$-3.26^{+0.02}_{-0.02}$&$-3.62^{+0.03}_{-0.04}$
 &$-5.30^{+4.20}_{-0.36}$  \\ 
WISEJ2102-44& T9& Regular&$0.501^{+0.001}_{-0.001}$
 &$0.422^{+0.001}_{-0.001}$&$0.29^{+0.01}_{-0.02}$
 &$0.36^{+0.01}_{-0.02}$&$-3.09^{+0.01}_{-0.02}$
 &$-2.96^{+0.01}_{-0.02}$&$-3.10^{+0.01}_{-0.02}$
 &$-9.03^{+1.77}_{-1.90}$  \\ 
WISEJ0359-54& Y0& Regular&$0.551^{+0.002}_{-0.002}$
 &$0.457^{+0.001}_{-0.001}$&$-0.02^{+0.03}_{-0.03}$
 &$0.02^{+0.03}_{-0.03}$&$-3.31^{+0.02}_{-0.03}$
 &$-3.23^{+0.02}_{-0.03}$&$-3.62^{+0.03}_{-0.04}$
 &$-4.32^{+0.05}_{-0.06}$  \\ 
WISEJ0734-71& Y0& Regular&$0.531^{+0.002}_{-0.002}$
 &$0.444^{+0.001}_{-0.001}$&$0.10^{+0.02}_{-0.02}$
 &$0.16^{+0.02}_{-0.02}$&$-3.23^{+0.02}_{-0.02}$
 &$-3.13^{+0.02}_{-0.02}$&$-3.41^{+0.02}_{-0.03}$
 &$-4.12^{+0.04}_{-0.04}$  \\ 
WISEJ1206+84& Y0& Regular&$0.535^{+0.002}_{-0.002}$
 &$0.447^{+0.001}_{-0.001}$&$0.27^{+0.02}_{-0.02}$
 &$0.32^{+0.02}_{-0.02}$&$-3.06^{+0.02}_{-0.02}$
 &$-2.97^{+0.02}_{-0.02}$&$-3.25^{+0.03}_{-0.03}$
 &$-3.92^{+0.04}_{-0.04}$  \\ 
WISEJ2209+27& Y0& Regular&$0.519^{+0.003}_{-0.003}$
 &$0.435^{+0.002}_{-0.002}$&$0.44^{+0.05}_{-0.05}$
 &$0.47^{+0.05}_{-0.05}$&$-3.24^{+0.02}_{-0.02}$
 &$-3.13^{+0.02}_{-0.02}$&$-3.36^{+0.03}_{-0.03}$
 &$-4.99^{+1.19}_{-0.19}$  \\ 
WISEPCJ2056+14& Y0& $YJH$-Pec.
 &$0.544^{+0.003}_{-0.003}$&$0.453^{+0.002}_{-0.002}$
 &$0.08^{+0.03}_{-0.03}$&$0.13^{+0.03}_{-0.03}$
 &$-3.22^{+0.03}_{-0.03}$&$-3.14^{+0.03}_{-0.03}$
 &$-3.48^{+0.04}_{-0.04}$&$-4.20^{+0.06}_{-0.06}$  \\ 
WISEJ0825+28& Y0.5& $Y$-Pec.
 &$0.536^{+0.003}_{-0.002}$&$0.447^{+0.002}_{-0.002}$
 &$0.63^{+0.05}_{-0.05}$&$0.64^{+0.05}_{-0.05}$
 &$-3.26^{+0.02}_{-0.02}$&$-3.17^{+0.02}_{-0.03}$
 &$-3.48^{+0.03}_{-0.03}$&$-8.98^{+2.05}_{-2.15}$  \\ 
WISEPCJ1405+55& Y0.5& Regular
 &$0.548^{+0.002}_{-0.002}$&$0.455^{+0.001}_{-0.001}$
 &$0.06^{+0.02}_{-0.02}$&$0.11^{+0.02}_{-0.02}$
 &$-3.23^{+0.02}_{-0.02}$&$-3.15^{+0.02}_{-0.02}$
 &$-3.52^{+0.02}_{-0.03}$&$-4.21^{+0.03}_{-0.04}$  \\ 
CWISEPJ1047+54& Y1& $Y$-Pec.
 &$0.518^{+0.001}_{-0.001}$&$0.435^{+0.001}_{-0.001}$
 &$0.61^{+0.01}_{-0.01}$&$0.68^{+0.01}_{-0.01}$
 &$-2.73^{+0.01}_{-0.01}$&$-2.62^{+0.01}_{-0.01}$
 &$-2.82^{+0.01}_{-0.01}$&$-4.35^{+0.11}_{-0.06}$  \\ 
CWISEPJ1446-23& Y1& $Y$-Pec.
 &$0.521^{+0.002}_{-0.002}$&$0.437^{+0.001}_{-0.001}$
 &$0.61^{+0.01}_{-0.01}$&$0.67^{+0.01}_{-0.01}$
 &$-2.76^{+0.01}_{-0.01}$&$-2.65^{+0.01}_{-0.01}$
 &$-2.87^{+0.01}_{-0.01}$&$-3.54^{+0.02}_{-0.02}$  \\ 
WISEAJ2354+02& Y1& $YJH$-Pec.
 &$0.546^{+0.002}_{-0.002}$&$0.454^{+0.001}_{-0.001}$
 &$0.18^{+0.02}_{-0.02}$&$0.23^{+0.02}_{-0.02}$
 &$-3.11^{+0.02}_{-0.02}$&$-3.03^{+0.02}_{-0.02}$
 &$-3.38^{+0.02}_{-0.02}$&$-4.26^{+0.04}_{-0.05}$  \\ 
WISEJ0535-75& Y1& $YJH$-Pec.
 &$0.543^{+0.003}_{-0.002}$&$0.452^{+0.002}_{-0.002}$
 &$0.21^{+0.05}_{-0.05}$&$0.23^{+0.05}_{-0.05}$
 &$-3.44^{+0.03}_{-0.03}$&$-3.36^{+0.03}_{-0.03}$
 &$-3.72^{+0.03}_{-0.04}$&$-8.17^{+2.42}_{-2.28}$  \\ 
WISEPAJ1541-22& Y1& $YJH$-Pec.
 &$0.524^{+0.003}_{-0.003}$&$0.439^{+0.002}_{-0.002}$
 &$0.68^{+0.05}_{-0.05}$&$0.70^{+0.05}_{-0.05}$
 &$-3.29^{+0.02}_{-0.03}$&$-3.19^{+0.03}_{-0.03}$
 &$-3.44^{+0.03}_{-0.03}$&$-8.55^{+2.30}_{-2.34}$  
\enddata

\tablenotetext{a}{These quantities incorporate corrections for oxygen sequestration, computed using the methodology described in \citet{Calamari_2024} and in \S\ref{metallicity}.}

\end{deluxetable*}

\begin{figure*}[htb!]
    \centering
    \includegraphics[width=\linewidth]{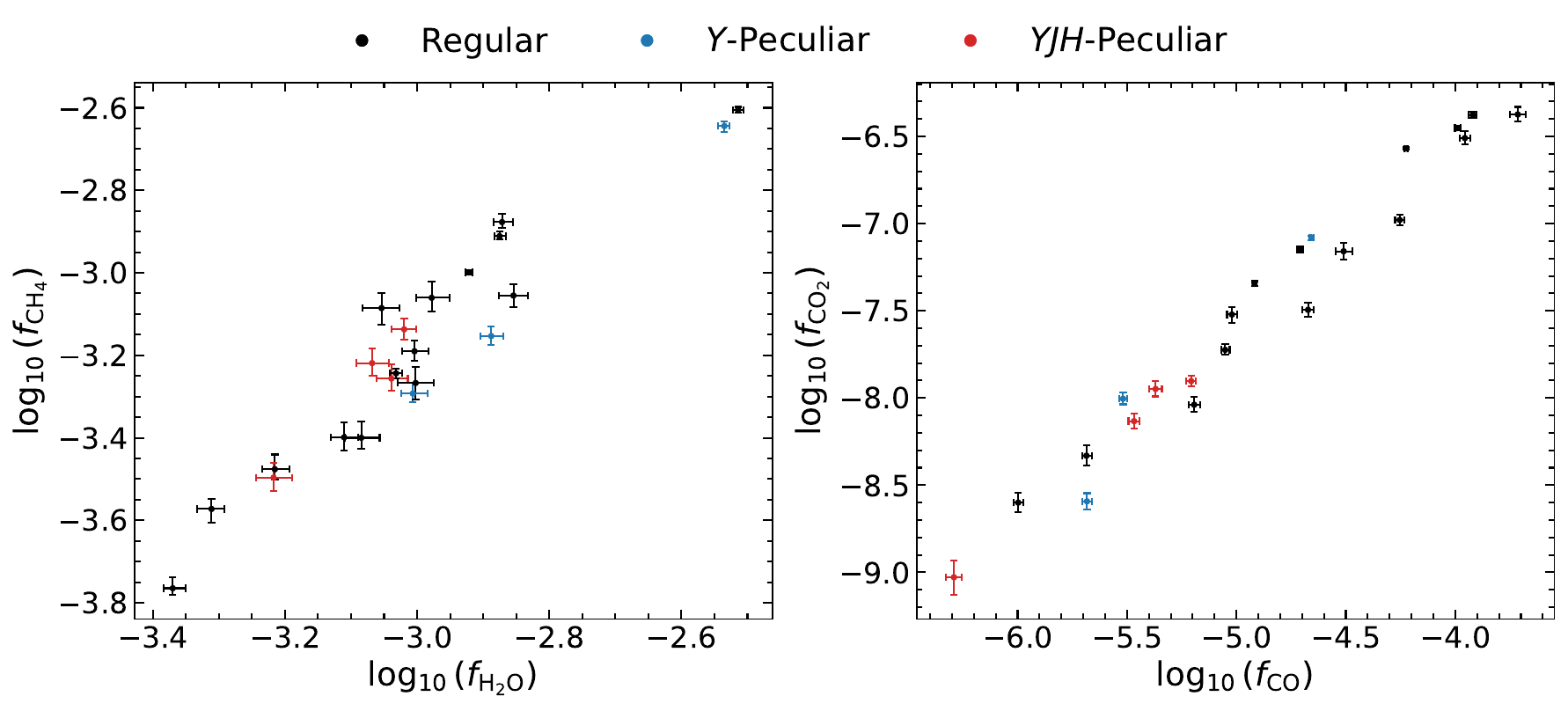}
    \caption{The left panel shows the trend between \fwater\ vs. \fmethane, while the right panel displays the trend between \fco\ vs. \fcoo in the sample. The error bars represent the 1$\sigma$ uncertainties on each measurement. The data points are color-coded by model spectral fit category: Regular (black), $Y$-Peculiar (blue), and $YJH$-Peculiar (red).}
    \label{h2o_ch4_co_co2}
\end{figure*}

One key advantage of atmospheric retrievals is their ability to measure individual gas mixing ratios directly from the spectra without making any a priori assumptions about the bulk atmospheric metallicity [M/H]$_\textrm{bulk}$ or (C/O)$_\textrm{bulk}$ ratio \citep{Line_2015, Zalesky_2019, Kothari_2024}. As mentioned in \S3, we are constraining the volume mixing ratios (VMRs) of nine gases: H$_2$O, CH$_4$, CO, CO$_2$, NH$_3$, H$_2$S, K, Na, and PH$_3$, while the rest of the atmosphere is assumed to be in the form of H$_2$ and He (see note c in Table \ref{priors}.). The retrieved VMRs for each object are summarized in Table \ref{chem_prop}.

From our analysis, we find that certain VMRs exhibit strong positive correlations. Notably, \fwater\ and \fmethane\ are both positively correlated with each other (see Figure \ref{h2o_ch4_co_co2}. [left]). Similarly, \fco\ and \fcoo\ are also positively correlated (Figure \ref{h2o_ch4_co_co2}. [right]).

Under equilibrium chemistry and at these cold temperatures, \fwater\ and \fmethane\ are expected to scale with the total oxygen and carbon abundance \citep{Lodders_2002, Marley_2021}, rather than being influenced by decoupled processes, i.e., their atmospheric chemistry is largely driven by local thermochemical equilibrium. The dominant decoupling process in cold atmospheres is typically vertical mixing, where material is dredged up from deeper and hotter layers to higher and cooler layers of the atmosphere. This vertical transport can “quench” the abundances of certain species (e.g., \co, \ammonia) at the level where the chemical interconversion timescale becomes longer than the mixing timescale, leading to chemical concentrations that are no longer directly tied to the local temperature or pressure. However, both \fwater\ and \fmethane\ have very short chemical timescales, allowing them to be dominated by thermochemical equilibrium even in the presence of vertical mixing \citep{Saumon_2003, Moses_2014}. The positive \fwater\ vs. \fmethane\ correlation observed in Figure~\ref{h2o_ch4_co_co2} [left] is therefore consistent with the expectation from equilibrium chemistry and suggests that neither species is strongly affected by disequilibrium chemistry in the objects studied.

Similarly, there is a positive correlation between \fco\ and \fcoo\ (Figure~\ref{h2o_ch4_co_co2}. [right]), which reflects their shared chemical pathways in the C$-$O network. Previous models of C$–$O kinetics (e.g., \citealt{Visscher_2006, Visscher_2010}) found that \coo\ primarily forms through reactions such as \co\ + OH $\rightleftharpoons$ \coo\ + H, linking its abundance closely to that of both \co\ and \water\ (the latter supplying OH radicals). In oxygen-rich, low-temperature environments like those of late-T and Y dwarf atmospheres, the \co\ + OH pathway links \coo\ directly to the availability of \co\ and \water\ \citep{Lodders_2002, Zahnle14}. Vertical mixing can further strengthen this link by transporting CO from deeper, hotter layers where the \co $\rightleftharpoons$\methane\ interconversion is shifted toward CO. Since \coo\ production is kinetically tied to CO and OH, the quenched CO abundance enhances \coo\ abundance relative to strict equilibrium predictions. Therefore, atmospheres with higher \fco, whether due to higher elemental abundances, higher temperatures, or disequilibrium chemistry due to vertical mixing, tend to also have higher \fcoo, provided sufficient \fwater\ is present \citep[see also][]{Beiler_2024, Wogan_2025}.

These correlations are a natural outcome of coupled chemistry and reinforce the internal consistency of retrievals \citep{Line_2015, Line_2017, Zalesky_2019}. Furthermore, they suggest that the dominant carbon- and oxygen-bearing species in brown dwarf atmospheres do not vary independently, but rather respond coherently to changes in fundamental atmospheric properties like elemental abundances.

\subsection{Bulk Atmospheric M/H \& C/O}
\label{metallicity}

\begin{figure}[htb!]
    \centering
    \includegraphics[width=\linewidth]{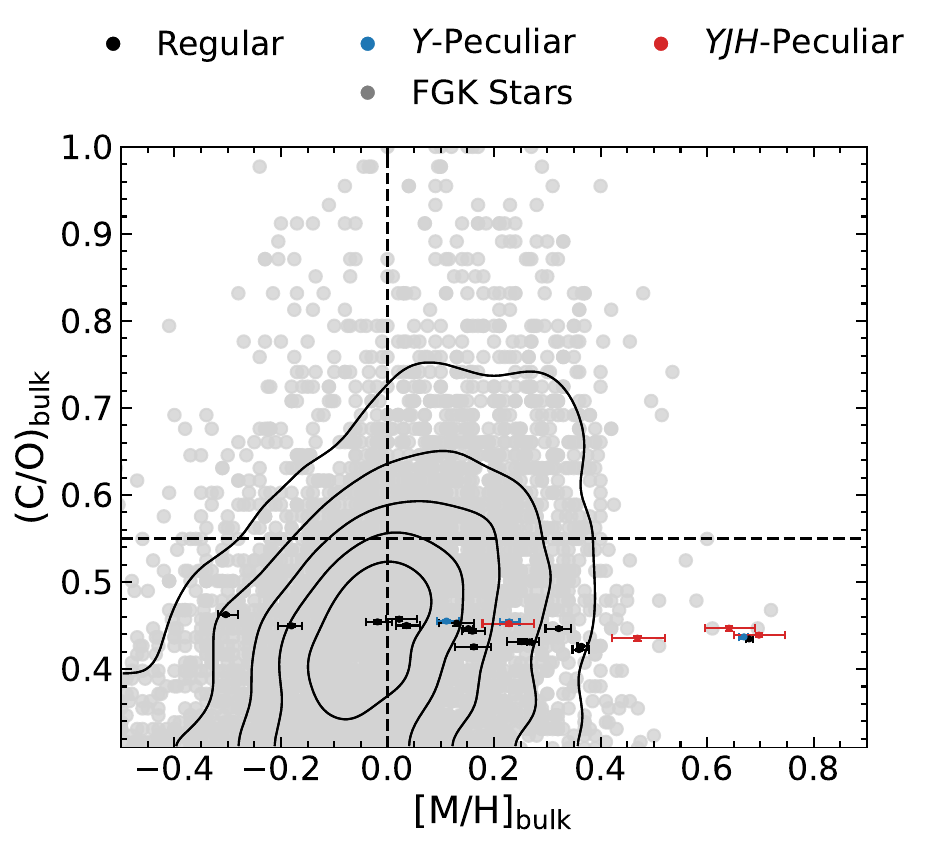}
    \caption{Calculated bulk atmospheric metallicity vs. C/O ratios for our sample, with 1$\sigma$ uncertainties derived from retrieved volume mixing ratios using equations \ref{M_H} and \ref{CO_ratio_eq}, respectively, which account for oxygen sequestration. The data points are color-coded by model spectral fit category: Regular (black), $Y$-Peculiar (blue), and $YJH$-Peculiar (red). The black dashed lines represent the solar metallicity and C/O ratio from \citet{Asplund_2009}. The grey points represent the FGK stars from within a 150 pc \citep{Hinkel_2014}, and the black contours show the smoothed density distribution of the FGK stars in the $(\mathrm{C/O})_{\mathrm{bulk}} $–$ [\mathrm{M/H}]_{\mathrm{bulk}}$ plane.}
    \label{M_H_C_O}
\end{figure}

\begin{figure*}[htb!]
    \centering
    \includegraphics[width=\linewidth]{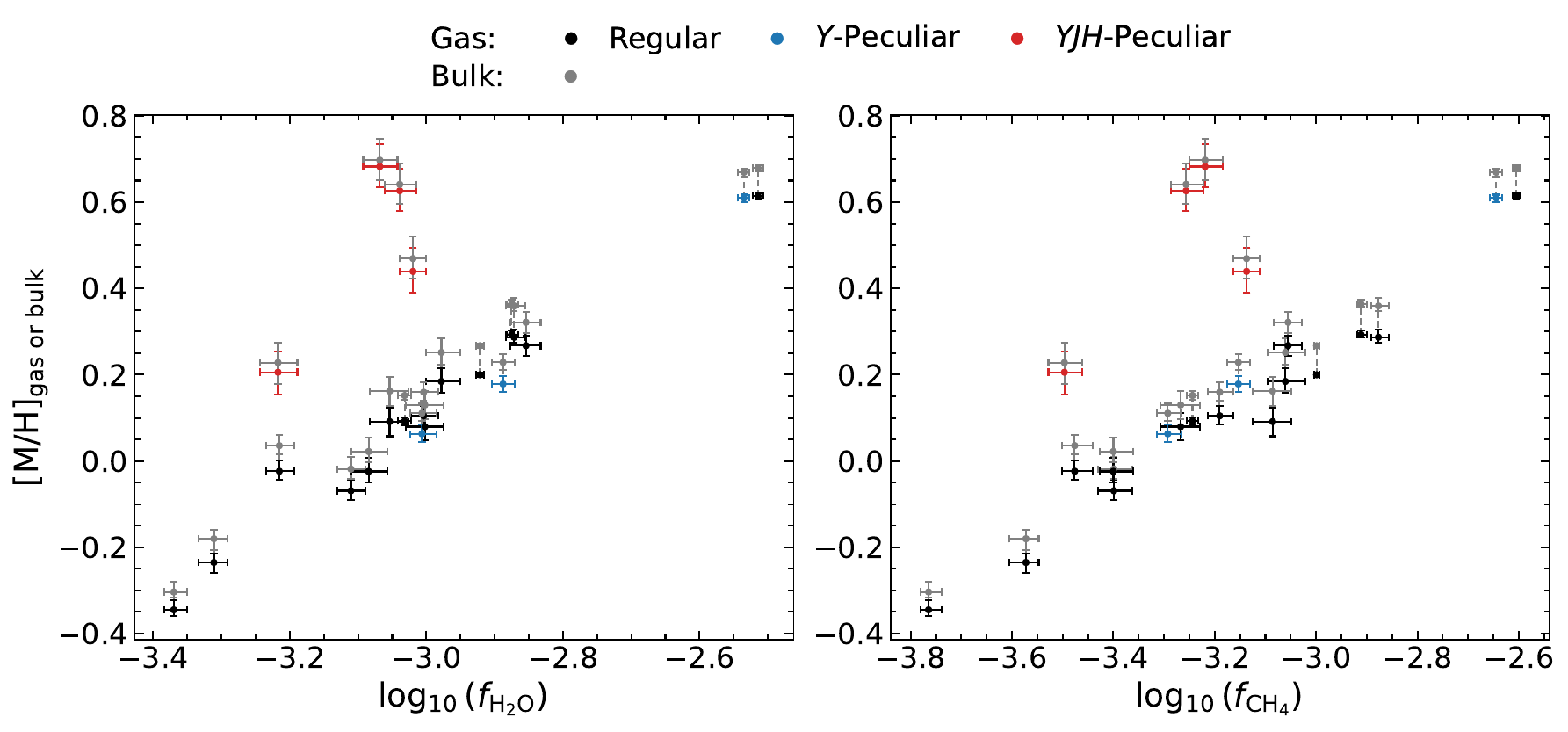}
    \caption{Trends of retrieved volume mixing ratios with gas-phase (non-grey) and bulk (grey) bulk metallicity. Left: \fwater\ vs. [M/H]$_\textrm{gas\,or\,bulk}$. Right: \fmethane\ vs. [M/H]$_\textrm{gas\,or\,bulk}$. The error bars represent the 1$\sigma$ uncertainties in both volume mixing ratio and metallicity. The data points are color-coded by model spectral fit category: Regular (black), $Y$-Peculiar (blue), and $YJH$-Peculiar (red).}
    \label{H2O_CH4_M_H}
\end{figure*}

The [M/H]$_\textrm{bulk}$ and the (C/O)$_\textrm{bulk}$ ratios are critical parameters for studying brown dwarfs as they tell us about the environment in which brown dwarfs were formed. These ratios also affect the substellar cooling rate which affects the observable properties of brown dwarfs (e.g. \citet{Burrows_2001}). Since our spectra covers all the dominant C- and O-bearing molecules within the atmosphere, we can accurately calculate the (C/O)$_\textrm{bulk}$ ratio of these cold brown dwarfs using the method laid out in \citet{Calamari_2024}. We first calculate the gas-phase (C/O)$_\textrm{gas}$ ratio using the retrieved mixing ratios as,

\begin{equation}
(\textrm{C}/\textrm{O})_\textrm{gas} = \frac{f_{\textrm{CO}} + f_{\textrm{CO}_{2}} + f_{\textrm{CH}_{4}}}{f_{\textrm{H}_{2}\textrm{O}} + f_{\textrm{CO}} + 2f_{\textrm{CO}_{2}}},  
\label{CO_ratio_eq}
\end{equation}

\noindent where we assume that carbon is primarily partitioned among CO, CO$_2$, and CH$_4$, and oxygen among H$_2$O, CO, and CO$_2$.

Condensation of refractory oxygen into silicate clouds (e.g., enstatite MgSiO$_3$ and forsterite Mg$_2$SiO$_4$; \citealt{Lodders_2002}) depletes oxygen from the gas-phase, increasing the above-cloud (i.e., observable) $(\textrm{C}/\textrm{O})_\textrm{gas}$ relative to the bulk value. \citet{Calamari_2024} derive an empirical relation (their Eq. 12) that accounts for this oxygen sink (typically $\sim$20\% in the solar neighborhood), which can be inverted to estimate the bulk ratio from an observed gas-phase value, assuming solar neighborhood abundances.  Treating $(\mathrm{C}/\mathrm{O})_{\mathrm{gas}}$ as the above-cloud C/O ratio, we derive the bulk C/O ratio as,

\begin{equation}
(\mathrm{C}/\mathrm{O})_{\mathrm{bulk}} 
\approx \frac{(\mathrm{C}/\mathrm{O})_{\mathrm{gas}}}
{1 + 0.371\,(\mathrm{C}/\mathrm{O})_{\mathrm{gas}}}.
\label{CO_ratio_bulk_eq}
\end{equation}

Since cold brown dwarf atmospheres are dominated by \water\ and \methane, which account for a majority of the observable oxygen and carbon budgets, respectively, we can calculate a robust estimate of the gas-phase metallicity (M/H)$_\textrm{gas}$ of these objects by summing the VMRs of the major observable metal-bearing species retrieved from our model using:

\begin{equation}
\begin{split}
\mathrm{(M/H)}_{\mathrm{gas}} &=
\frac{
\mathrm{O_{\mathrm{gas}}} + \mathrm{C_{\mathrm{gas}}} + \mathrm{N} + \mathrm{S} + \mathrm{P} + \mathrm{Na} + \mathrm{K}}%
{2\left([1-\sum f_\textrm{metals}] \times 0.84\right)+\sum n f_{{\rm H}_n}} \\[6pt]
&=
\frac{ f_{\textrm{H}_{2}\textrm{O}} + f_{\textrm{CH}_{4}} + 2f_{\textrm{CO}} + 3f_{\textrm{CO}_{2}} +}{%
  2\left([1-\sum f_\textrm{metals}] \times 0.84\right)+}... \\[6pt]
&\quad \frac{f_{\textrm{NH}_{3}} + f_{\textrm{H}_{2}\textrm{S}} + f_{\textrm{K}}
   + f_{\textrm{Na}} + f_{\textrm{PH}_{3}}}
   {\sum n f_{{\rm H}_n}},
   % {+\, 4f_{\mathrm{CH_4}} + 3f_{\mathrm{NH_3}}
   % + 2f_{\mathrm{H_2O}} + 2f_{\mathrm{H_2S}} + 3f_{\mathrm{PH_3}} } .
\end{split}
\label{M_H_atm_wq}
\end{equation}

\noindent where the numerator sums the VMRs of metal-bearing species, with coefficients accounting for the number of metal atoms per molecule (e.g., 2 for \co, 3 for \coo). The denominator calculates the total hydrogen content by assuming that the VMR of non-metal gas fraction (1$-$$\sum f_\textrm{metals}$) consists of H$_2$ and He, where H$_2$ represents 84\% of this fraction (hence the factor of 0.84). The factor of 2 converts the H$_2$ VMR into atomic H equivalent and $n f_{{\rm H}n}$ represents the hydrogen contribution from molecules containing $n$H atoms (e.g., $2f_{\mathrm{H_2O}}$, $4f_{\mathrm{CH_4}}$, etc.) Finally, the gas-phase metallicity relative to solar value ((M/H$_\odot$) $\approx$ 8.42 $\times$ 10$^{-4}$; \citet{Asplund_2009}) is given by:

\begin{equation}
[\textrm{M}/\textrm{H}]_\mathrm{gas} = \log_{10} \left ( \frac{(\textrm{M}/\textrm{H})_\textrm{gas}}{(\textrm{M}/\textrm{H})_\odot} \right ).  
\label{M_H}
\end{equation}

However, $[\textrm{M/H}]_{\mathrm{gas}}$ does not account for oxygen removed from the gas-phase by condensation into silicate clouds. Therefore, we adapt \citet{Calamari_2024} methodology to account for the fraction of the total oxygen inventory sequestered into condensates as:

\begin{equation}
\mathrm{O}_{\mathrm{sink}} \approx  0.371\times(\mathrm{C}/\mathrm{O})_\mathrm{bulk},
\label{CO_sink}
\end{equation}

\noindent where $\mathrm{O}_{\mathrm{sink}}$ is the fraction of bulk oxygen in condensates in a given atmosphere. Using equations~\ref{CO_ratio_bulk_eq} and \ref{CO_sink}, together with equation~10 from \citet{Calamari_2024}\footnote{In equation~10 of \citet{Calamari_2024}, the carbon content refers to the gas-phase carbon content. When expressing the bulk C/O ratio in terms of gas-phase quantities, the carbon term appears in both numerator and denominator and therefore cancels algebraically.}, we can express the bulk oxygen abundance as,

\begin{equation}
\begin{split}
\mathrm{O}_{\mathrm{bulk}} 
&= \mathrm{O}_{\mathrm{gas}} \times (1 - \mathrm{O}_{\mathrm{sink})}
 \\
&\approx \mathrm{O}_{\mathrm{gas}} \times 0.371 \times (\mathrm{C}/\mathrm{O})_{\mathrm{bulk}} .
\end{split}
\end{equation}

Using  $\mathrm{O}_{\mathrm{bulk}}$, we can calculate the bulk atmospheric metallicity as,  

\begin{equation}
\mathrm{(M/H)}_{\mathrm{bulk}} =
\frac{
\mathrm{O_{\mathrm{bulk}}} + \mathrm{C_{\mathrm{gas}}} + \mathrm{N} + \mathrm{S} + \mathrm{P} + \mathrm{Na} + \mathrm{K}}{2\left([1-\sum f_\textrm{metals}] \times 0.84\right)+\sum n f_{{\rm H}_n}}.
\end{equation}

\noindent Here, each elemental abundance is the sum of the VMRs of the corresponding element-bearing species (e.g. $\mathrm{C_{\mathrm{gas}}}$ = $f_\textrm{\methane}$+$f_\textrm{\co}$+$f_\textrm{\coo}$). We than normalize the bulk metallicity relative to solar value ((M/H$_\odot$) $\approx$ 8.42 $\times$ 10$^{-4}$ \citet{Asplund_2009}) as,
\begin{equation}
[\textrm{M/H}]_\textrm{bulk} = 
\log_{10}\left(\frac{\mathrm{(M/H)}_{\mathrm{bulk}}}{(\mathrm{M/H})_{\odot}}\right).
\label{M_H_bulk}
\end{equation}

Table \ref{chem_ratio} lists the calculated gas-phase (C/O)$_\textrm{gas}$ and [M/H]$_\textrm{gas}$, as well as  the corresponding bulk (C/O)$_\textrm{bulk}$ and [M/H]$_\textrm{bulk}$ for all the objects in our sample. Figure \ref{M_H_C_O} shows the relationship between the calculated bulk atmospheric metallicity and C/O ratios for our sample, overlaid on measurements of FGK stars from within a 150 pc \citep{Hinkel_2014}. We find that although the bulk C/O ratios of our sample objects are sub-solar and most exhibit super-solar metallicities, they fall within the compositional spread of the local FGK stellar population, indicating no significant deviation from nearby stellar abundance trends.

Finally, comparing the calculated [M/H]$_\textrm{bulk}$ with the retrieved VMRs we find that \fwater \ and \fmethane \ are positively correlated with [M/H]$_\textrm{bulk}$ across our sample (see Figure \ref{H2O_CH4_M_H}.). This trend suggests that both species are jointly tracing the overall metal content ([M/H]$_\textrm{bulk}$) of the atmosphere. Since water and methane are the dominant oxygen (O)- and carbon (C)-bearing molecules, respectively, their abundances are sensitive to the elemental abundances of O and C (as mentioned in \S\ref{mixingratios}), which are set by the object's bulk metallicity. The observed correlation indicates that as metallicity increases, the availability of O and C increases proportionally, leading to enhanced formation of both \water \ and \methane. Conversely, atmospheres with lower [M/H]$_\textrm{bulk}$ show a simultaneous depletion in both molecules. This consistency supports the interpretation that retrieved \water\ and \methane\ chemistry is primarily governed by elemental abundances under near-equilibrium conditions, and that our spectra are effectively capturing the signatures of overall atmospheric composition.

\subsection{Atmospheric Sulfur}

 In cool brown dwarf atmospheres, equilibrium chemistry predicts that sulfur resides primarily in \hhs, making it a reasonable proxy for the total sulfur content \citep{Visscher_2006}. Although our dataset does not explicitly exhibit \hhs \ spectral features, atmospheric retrievals reveal a striking diversity in \hhs \ mixing ratios across the sample, with $\log_{10}$\fhhs\ values ranging from $-$9.47 to $-$3.29. The hotter objects (spectral type T) tend to have $\log_{10}$\fhhs\ values that are near the lower bound of the prior and are unconstrained, whereas colder objects (spectral type Y) show higher and constrained $\log_{10}$\fhhs\ values, except for the $YJH$-peculiar object WISE 0825+28, whose \fhhs\ remains unconstrained.

We calculate the gas-phase sulfur-to-hydrogen ratio (S/H)$_\textrm{gas}$ using:

\begin{equation}
(\textrm{S}/\textrm{H})_\textrm{gas} =
\frac{f_{\mathrm{H_2S}}}{%
2\left([1-\sum f_\textrm{metals}] \times 0.84\right)+\sum n f_{{\rm H}_n}}
\label{S_H_atm}
\end{equation}

\noindent where the numerator accounts for sulfide-bearing species, which in our case is assumed to be exclusively in the form of \hhs. The denominator represents the total atmospheric hydrogen content, as described in \S4.4. For comparison, the solar sulfur-to-hydrogen ratio is $(\textrm{S}/\textrm{H})_\odot = 10^{-4.88}$ \citep{Asplund_2009}.

We find that 10 of the 22 brown dwarfs exhibit (S/H)$_\textrm{gas}$ values greater than the solar ratio.  Kinetic/transport processes (e.g., vertical mixing) may modulate how efficiently sulfur is removed from the gas-phase and such processes would  act to preserve sulfur rather than produce supersolar S/H values. In an atmospheric ``rainout’’ scenario, abundant refractory sulfides such as FeS and MgS are unlikely to form because Fe and Mg are sequestered deeper into the atmosphere at higher temperatures; thus \hhs\ is expected to remain the dominant sulfur carrier, with possible minor depletion into condensates such as Na$_2$S, MnS, or ZnS at$\sim$700$–$1300 K \citep{Lodders_2002, Visscher_2006}. At very low temperatures, molecular condensates such as NH$_4$SH or solid H$_2$S may become more efficient at removing sulfur from the vapor phase. Thus, large variations in \fhhs\ are not straightforward to explain, especially in cold atmospheric conditions.

For the remaining objects, (S/H)$_\textrm{gas}$ lies below the solar ratio, consistent with some level of sulfur sequestration into condensates at low temperatures (500–1000 K) \citep{Morley_2012, Gao_2021}. Disequilibrium processes such as vertical mixing could further modulate \hhs\ abundances, particularly in warmer T dwarfs, where \hhs\ might be chemically converted or transported to deeper, less observable layers \citep{Gao_2021}.

Overall, our retrievals underscore sulfur’s diagnostic value as a tracer of both condensation chemistry and formation environments. Because \hhs\ remains the overwhelmingly dominant sulfur-bearing species in these conditions, (S/H)$_\textrm{gas}$ provides a relatively direct measurement of the atmospheric sulfur abundance, complementing metallicity and C/N/O ratios as a probe of brown dwarf chemistry \citep{Marley_2021, Molli_re_2022}.

\subsection{Atmospheric Phosphine}

\begin{figure*}[htb!]
    \centering
    \includegraphics[width=\linewidth]{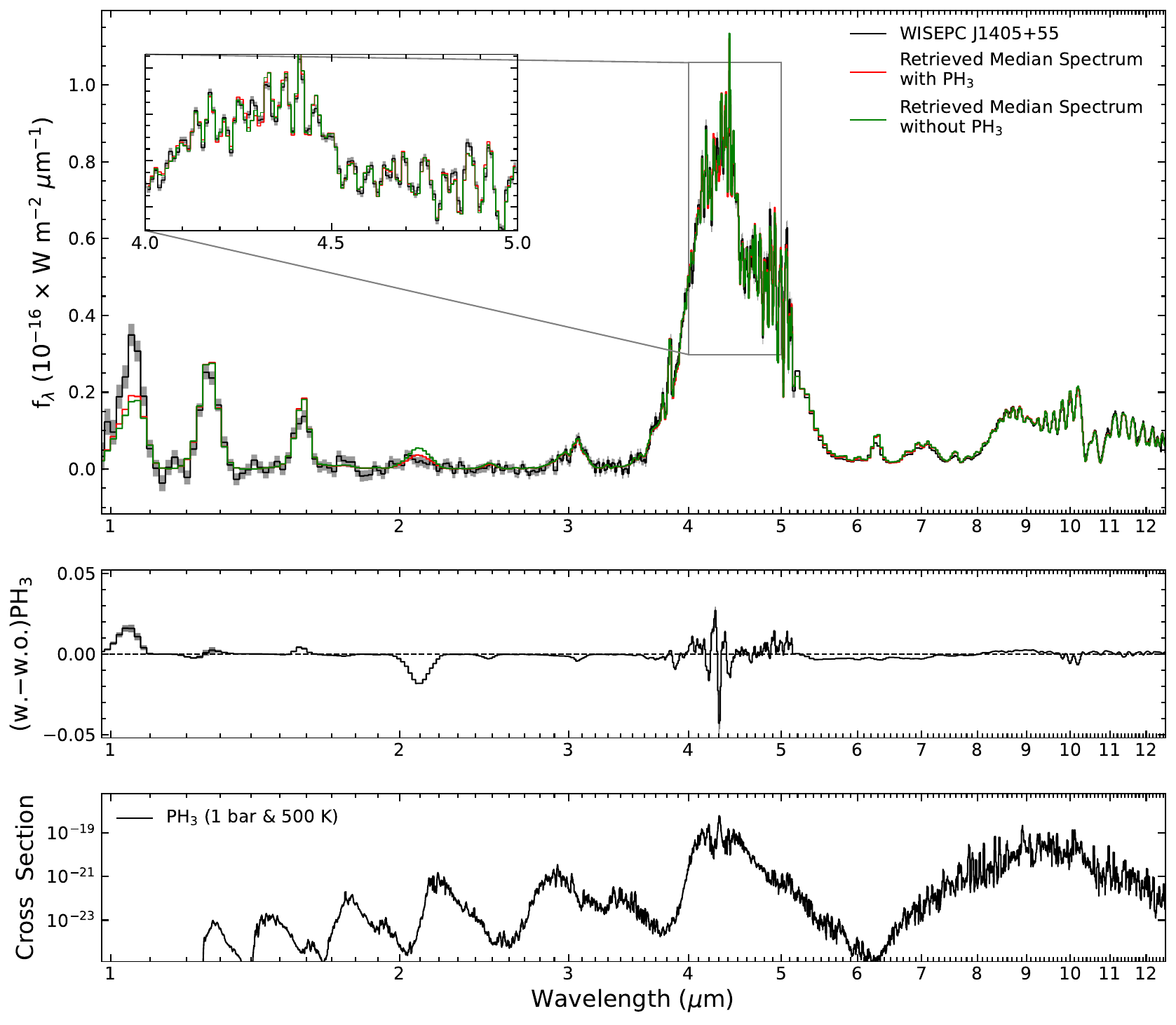}
    \caption{The top panel shows the observed JWST spectrum of WISEPC J1405+55 (spectral type Y0.5) in black covering $\sim$1–12.2 $\mu$m in f$_{\lambda}$  with 1$\sigma$ uncertainties in grey. The red and green line represent the retrieved median spectrum with and without PH$_{3}$ in the retrieval model, receptively, with a zoomed in version of the 4 to 5 \mum\ region. The middle panel shows the difference between the two retrieved models (with and without PH$_{3}$) with 1$\sigma$ credible intervals in grey. The bottom panel shows the PH$_{3}$ cross-sections at 500 K and 1 bar.}
    \label{1405_ph3_compare_2}
\end{figure*}

\begin{deluxetable*}{lcccccc}[htb!]
\centering
\tablecaption{Retrieved Phosphine Model Comparison \label{ph3_comparison}}
\tablehead{
\colhead{Object Name} &
\colhead{SpT} &
\colhead{Pecularity} &
\colhead{$\log_{10}(f_\textrm{PH$_{3}$})$} &
\colhead{$\Delta$AIC} \tablenotemark{a} &
\colhead{$\Delta$BPICs} \tablenotemark{a} &
\colhead{Significance} \tablenotemark{b}
}
\startdata
WISEPC J1405+55   & Y0.5 & $Y$-Pec.& $-8.10^{+0.08}_{-0.08}$ & -77.71 & -76.71 & ``Major Discoveries"\\
WISEPA J0313+78   & T8.5 & Regular& $-7.29^{+0.07}_{-0.07}$ & -46.90 & -47.68 & ``Major Discoveries"\\
ULAS J1029+09     & T8   & Regular& $-6.79^{+0.05}_{-0.05}$ & -42.95 & -46.19 & ``Major Discoveries"\\
WISEA J2102$-$44  & T9   & Regular& $-7.06^{+0.05}_{-0.05}$ & -38.03 & -36.46 & ``Major Discoveries"\\
WISEA J2159$-$48  & T9   & Regular& $-7.59^{+0.08}_{-0.08}$ & -31.12 & -30.81 & ``Major Discoveries"\\
WISEPA J1959$-$33 & T8   & Regular& $-7.01^{+0.09}_{-0.09}$ & -29.80 & -30.01 & ``Major Discoveries"\\
WISE J1206+84     & Y0   & Regular& $-7.52^{+0.08}_{-0.09}$ & -28.87 & -28.70 & ``Major Discoveries"\\
WISE J0734$-$71   & Y0   & Regular& $-7.79^{+0.09}_{-0.09}$ & -20.53 & -20.73 & ``Detections"\\
WISEPA J2354$-$33 & Y1   & $Y$-Pec.& $-8.27^{+0.15}_{-0.15}$ &  -6.75 &  -7.88 & ``Minor Results\\
WISE J0359$-$54   & Y0   & Regular& $-8.65^{+0.37}_{-0.37}$ &  -0.71 &  -2.66 & ``Indications"
\enddata

\tablenotetext{a}{The $\Delta$ represents the difference between the retrieval model with and without PH$_{3}$.}

\tablenotetext{b}{The qualitative significance of the PH$_3$ detection based on the relative strength of model preference as suggested by \citet{thorngren_2025}}

\end{deluxetable*}

PH$_3$ is a key tracer of non-equilibrium chemistry in substellar atmospheres, providing insight into vertical mixing and the quenching of phosphorus-bearing species \citep{Visscher_2006, Visscher_2010}. Thermochemical equilibrium predicts observable PH$_3$ in cool brown dwarfs (T$<$1000 K) when vertical transport dredges PH$_3$ from hotter layers faster than it is oxidized (e.g., to P$_4$O$_6$) \citep{Fegley_1994, Visscher_2006}. Although PH$_3$ is prominent in the spectrum of Jupiter, especially between 4.3 to 4.6 \mum\ region \citep{Larson_1977, FLETCHER_2009}, PH$_3$ detections in brown dwarfs remain limited to WISE 0855$-$07 with a VMR of 10$^{-9.24\pm0.07}$ \citep{Rowland_2024} and Wolf 1130C with a VMR of 10$^{-7.00\pm0.04}$ \citep{Burgasser_2025}.

In our analysis, we retrieve the VMR for PH$_3$ and constrain it in 15 brown dwarfs in our sample, with the $\log_{10}$(VMR) values ranging from $\sim$$-$6.8 to $-$8.7. To verify the robustness of the retrieved PH$_3$ VMR values, we ran another set of retrievals for these 15 brown dwarfs without PH$_3$ in the retrieval model spectra and compared the two sets of model spectra. We find that the residual between the retrieved model spectra generated with and without PH$_3$ exhibit features that track PH$_3$ cross-sections, especially between 4 and 5 \mum\ (shown in Figure \ref{1405_ph3_compare_2}), which has a negative feature centered on the Q-branch near $\sim$4.3 \mum, with two adjoining negative structure at $\sim$4.25 \mum\ (R-branch) and $\sim$4.4 \mum\ (P-branch). To quantify the model preference we employ AIC \citep[Akaike Information Criterion;][]{AKAIKE_1979} and BPICs \citep[Bayesian Predictive Information Criterion;][]{Ando_2011}, as recommended by \citet{thorngren_2025}. This quantification (Table \ref{ph3_comparison}) shows that 10 objects prefer models with PH$_3$ ($\Delta$AIC \&  $\Delta$BPIC $<$ 0). Among these 10 objects, seven meet the ``Major Discovery" threshold ($\Delta$AIC and $\Delta$BPIC $\lesssim-28$) as laid out by \citet{thorngren_2025}. One qualifies as a ``Detection", and the remaining two objects qualify as ``Minor Results" and ``Indications", respectively. 

These PH$_3$ detections are largely driven by the feature between 4 to 5 \mum\ window as seen in the residual (Figure \ref{1405_ph3_compare_2}), where the R-branch (4.20–4.28~\,$\mu$m), Q-branch (4.28–4.33\,$\mu$m), and P-branch (4.33–4.50\,$\mu$m) dominate the cross-section (Fig.~\ref{1405_ph3_compare_2}). These features are consistently reproduced across the 10 objects to a varying degree with constraints for the PH$_3$ VMR. The objects with the strongest significance for PH$_3$ detection exhibit residuals outside of the 4 to 5 \mum\ regions (e.g. $\sim$2 to 3 \mum), where the PH$_3$ cross-section blends with H$_2$O and CH$_4$ cross-sections. Therefore, the 4 to 5 \mum\ band provides the cleanest and most diagnostic evidence for PH$_3$ in our sample, mirroring Jupiter’s phosphine signature in the same spectral range.

While the analysis above strongly suggests we have detected PH$_3$, we continue to refer to them as ``tentative" because of the lack of a clear spectroscopic signature. Overall, these results indicate that PH$_3$ is tentatively present at measurable levels in a subset of late-T and Y dwarfs based on: (1) structured 4–5 \mum\ residuals aligned with the PH$_3$ band and (2) information-criterion support for PH$_3$-inclusive model preference.

\subsection{\ammonia, Na, \& K}
\label{mixingratios_rest}

Across the sample, the retrieved \ammonia\ abundance broadly follows the expectation from thermochemical equilibrium predictions, i.e., as the objects cool from T6--T8, \ammonia\ becomes more abundant as nitrogen favors NH$_3$ at lower temperatures (e.g., \citealt{Lodders_2002}). However, below T8, \ammonia\ VMR clusters within an order of magnitude ($\sim$10$^{-5}$ to 10$^{-4}$) with no systematic trend in NH$_3$'s VMR. Our range of \ammonia\ VMR values also fall within the reported values from \citet{Zalesky_2019}, where the retrieved \ammonia\ VMR for late-T and Y dwarfs fell between 10$^{-5}$ to 10$^{-4}$. We note, however, that in \citet{Zalesky_2019} the spectra did not exhibit clear spectroscopic NH$_3$ absorption features.

We include Na and K in our retrieval framework because the pressure-broadened wings of their resonance doublets significantly influence the Y-band continuum, even though the line cores (located at optical wavelengths) lie outside our spectral coverage. In late-T and Y dwarfs, the far wings of the Na and K lines can suppress flux shortward of the $J$$-$band; however, the quantum mechanical treatment of these line profiles under high-pressure H$_2$/He conditions remains challenging. As a result, the retrieved Na and K VMRs are model-dependent and should be interpreted with caution.

We find no clear systematic trend in K VMR across the sample. For Na, however, the hotter objects exhibit larger retrieved abundances than the colder Y dwarfs. This qualitative behavior is consistent with equilibrium condensation chemistry, in which Na and K are progressively removed from the gas phase at lower temperatures through the formation of Na$_2$S and KCl condensates. However, we would like to note that Na and K VMRs could also be partially compensating for mismatches in the modeled $Y$$-$band opacity, and therefore their VMRs should be interpreted with caution.

\subsection{Physical properties: $M$, $R$, $g$, and $T_\textrm{eff}$}
\label{sec:physicalproperties}

The mass ($M$) and radius ($R$) of each object are key parameters in determining the object's evolutionary stage, and are retrieved as part of our atmospheric model. The retrieved masses and radii (see Table \ref{phys_prop}) span from $\sim$6 to 77 $\mathcal{M}^\textrm{N}_\textrm{Jup}$ and $\sim$0.66 to 1.53 $\mathcal{R}^\textrm{N}_\textrm{Jup}$, respectively. The radii range is consistent with theoretical predictions that brown dwarf radii are largely constant across a broad mass range due to competing electron-degeneracy pressure and Coulomb effect \citep{Burrows_1993, Jagadeesh_2026}. While the mass range is reasonable for Y dwarfs, which are expected to be predominantly low-mass and/or young objects \citep[e.g.,][]{Kirkpatrick_2012}, a notable discrepancy arises for warmer objects. For sources with \teff $\gtrsim 600$ K (spectral types of approximately T8 and earlier), the retrievals systematically predict higher masses than expected from evolutionary models \citep{Saumon_2008, Morley_2012}, except for three objects: SDSS J1624+00 (T6), WISE J1501-40 (T6), and SDSSp J1346-00 (T6.5). This is a known effect where atmospheric retrievals infer a higher mass than forward models for a given effective temperature \citep[e.g.,][]{Line_2015, Zalesky_2019}. The reason for this discrepancy is still an active area of research, but it may be due to a difference in the retrieved and forward model thermal profiles.

\noindent Using the retrieved mass and radius, we calculated the surface gravity as,

\begin{equation}
g = \frac{GM}{R^2},
\label{gravity_eq}
\end{equation}

\noindent where $G$ is the gravitational constant with a value of 6.67430 $\times 10^{-11} \textrm{ m}^3 \textrm{ kg}^{-1} \textrm{ s}^{-2}$\citep[]{Mamajek_2015}. The resulting \logg \  values range from $\sim$4 to 5.5[cm s$^{-2}$], capturing both low-gravity (young or low-mass) and high-gravity (older or more massive) brown dwarfs in the sample. 

Using the retrieved radius ($R$), we also calculated \teff \ using the Stefan-Boltzman law,

\begin{equation}
T_\mathrm{eff} = \left ( {\frac{L_\mathrm{bol}}{4 \pi \sigma R^2}}\right) ^{\frac{1}{4}},
\label{teff_eq}
\end{equation}

\noindent where $L_\mathrm{bol}$ is the bolometric luminosity given by $L_\mathrm{bol} = 4 \pi d^2 F_\mathrm{bol}$, $d$ is the retrieved distance, and $F_\textrm{bol}$ is the bolometric flux calculated by integrating the model spectrum over all wavelengths (0 to $\infty$). To account for light emerging at wavelengths shorter than the minimum wavelength and longer than the maximum wavelength of a given model spectrum, we linearly interpolated the model from zero flux at zero wavelength to the flux at minimum wavelength, and then extended the model to $\lambda=\infty$ using a Rayleigh-Jeans tail, where the flux densities for Rayleigh-Jeans tail ($f_{\lambda,\textrm{RJ}}$) is proportional to $\lambda^{-4}$; the constant of proportionality is calculated using the flux density of the last model wavelength.   

We compare our calculated $T_\mathrm{eff}$ values with the reported $T_\mathrm{eff}^{\mathrm{uni}}$ from \citet{Beiler_2024_b}, as listed in Table \ref{phys_prop}. Overall, our derived $T_\mathrm{eff}$ values are broadly consistent with the reported $T_\mathrm{eff}^{\mathrm{uni}}$. The discrepancies between the two primarily arise from the assumption in \citet{Beiler_2024_b} of a fixed radius of $\sim1\pm0.1~\mathcal{R}^\mathrm{N}_\mathrm{Jup}$, whereas our median retrieved radii span a wider range of $0.66$–$1.53~\mathcal{R}^\mathrm{N}_\mathrm{Jup}$. Additional differences in the effective temperature values also arise from the mismatches between the model and observed flux densities in the near-infrared region of the spectrum for $Y$ and $YJH$-peculiar objects.

To assess how our derived $\log_{10}(g)$ and $T_\mathrm{eff}$ compare to theoretical predictions, we place our sample on the cloudless Sonora-Bobcat evolutionary models \citep{Marley_2021} (see Figure \ref{evolution}). These evolutionary models predict how $\log_{10}(g)$ and $T_\mathrm{eff}$ evolve over time for a range of masses. 

We find that most of our sample falls within the expected region for field-age late-T and Y dwarfs, typically assumed to have $\log_{10}(g) \sim 5$ [cm s$^{-2}$] \citep{Saumon_2008}, with an upper bound of 5.3 [cm s$^{-2}$]. However, over half of our objects have a derived $\log_{10}g < 5$ [cm s$^{-2}$], which puts them below the typically expected $\log_{10}(g)$ of $\sim 5$ [cm s$^{-2}$], implying that a substantial fraction of our sample may include younger than typically expected age for an old field population. While most of the objects fall within the generally assumed age range of 1 to 10 Gyr, four objects (ULAS J1029+09, WISE J0247+37,  WISE J2102$-$44, and CWISEP J1047+54) appear to have ages slightly exceeding 10 Gyr and one object (WISEPC J2056+14) having a lower age than 1 Gyr based on their derived surface gravity.

To evaluate for potential memberships of each object in our sample to known nearby young associations, we applied the BANYAN $\Sigma$, a Bayesian analysis framework \citep{Gagne_2018}, which compares an object’s sky position, proper motion, parallax, and radial velocity to the kinematic models of the known young moving groups within 150 pc. Using the coordinates, proper motion, and parallax for the objects in our sample, we find that all our objects have a high probability of belonging to the field population, with little to no likelihood of association with any of the known moving groups.

\begin{figure*}[htb!]
    \centering
    \includegraphics[width=\linewidth]{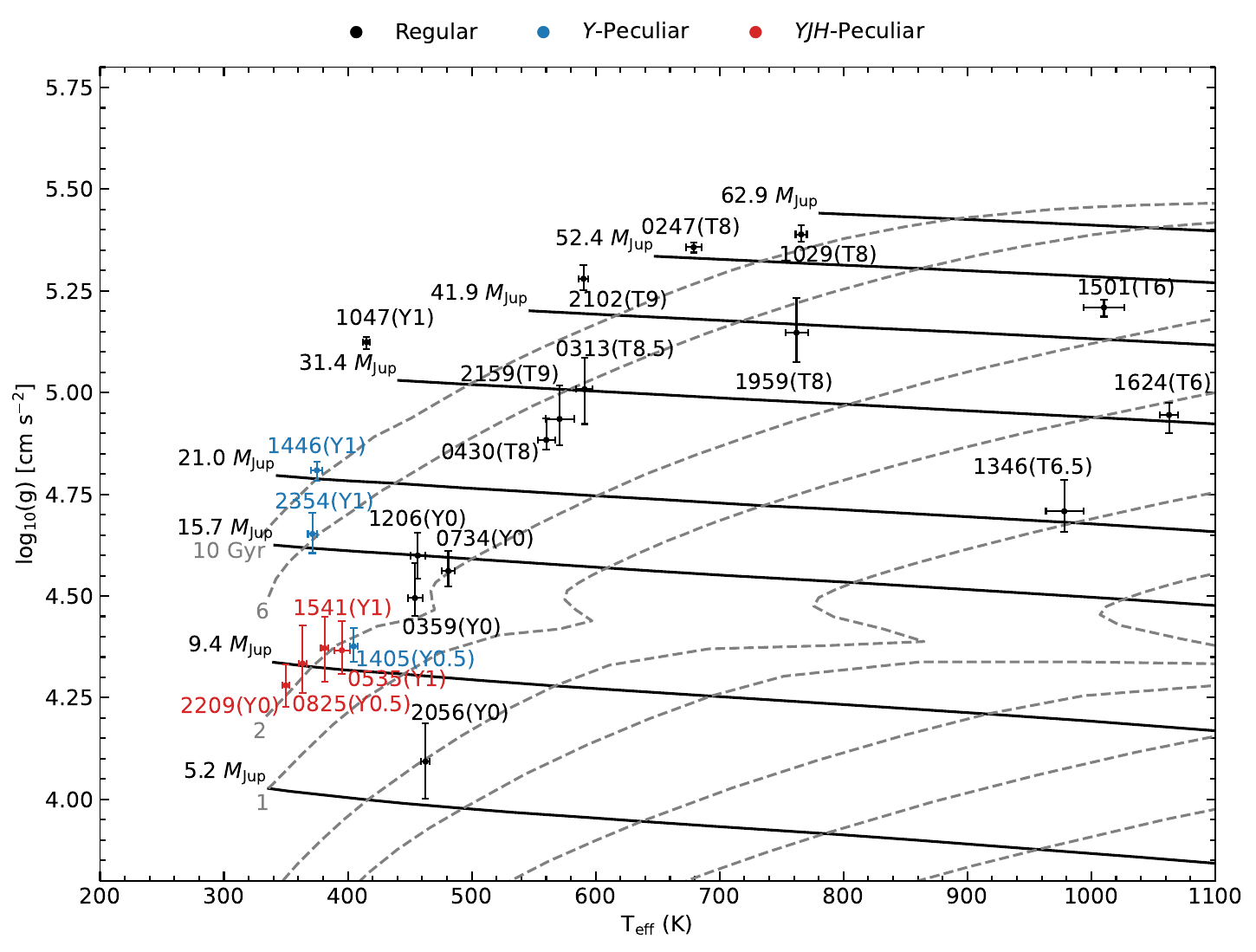}
    \caption{Evolution of Bobcat Sonora solar metallicity cloudless brown dwarfs in the effective temperature surface gravity plane \citep{Marley_2021}. The black lines are cooling tracks for brown dwarfs with masses of 62.9, 52.4, 41.9, 31.4, 21, 15.7, 9.4, and 5.2 $\mathcal{M}^\textrm{N}_\textrm{Jup}$, while the gray curves are isochrones for ages of 10, 6, 2, 1, .4, .2, .08, .04, and .02 Gyr. Overplotted are the calculated \teff\ and \logg\ values with their respective 1$\sigma$ uncertainties for the entire sample. The data points are color-coded by model spectral fit category: Regular (black), $Y$-Peculiar (blue), and $YJH$-Peculiar (red).}
    \label{evolution}
\end{figure*}

\subsection{Comparison to Lueber et al. (2026)}

\begin{figure*}[htb!]
    \centering
    \includegraphics[width=\linewidth]{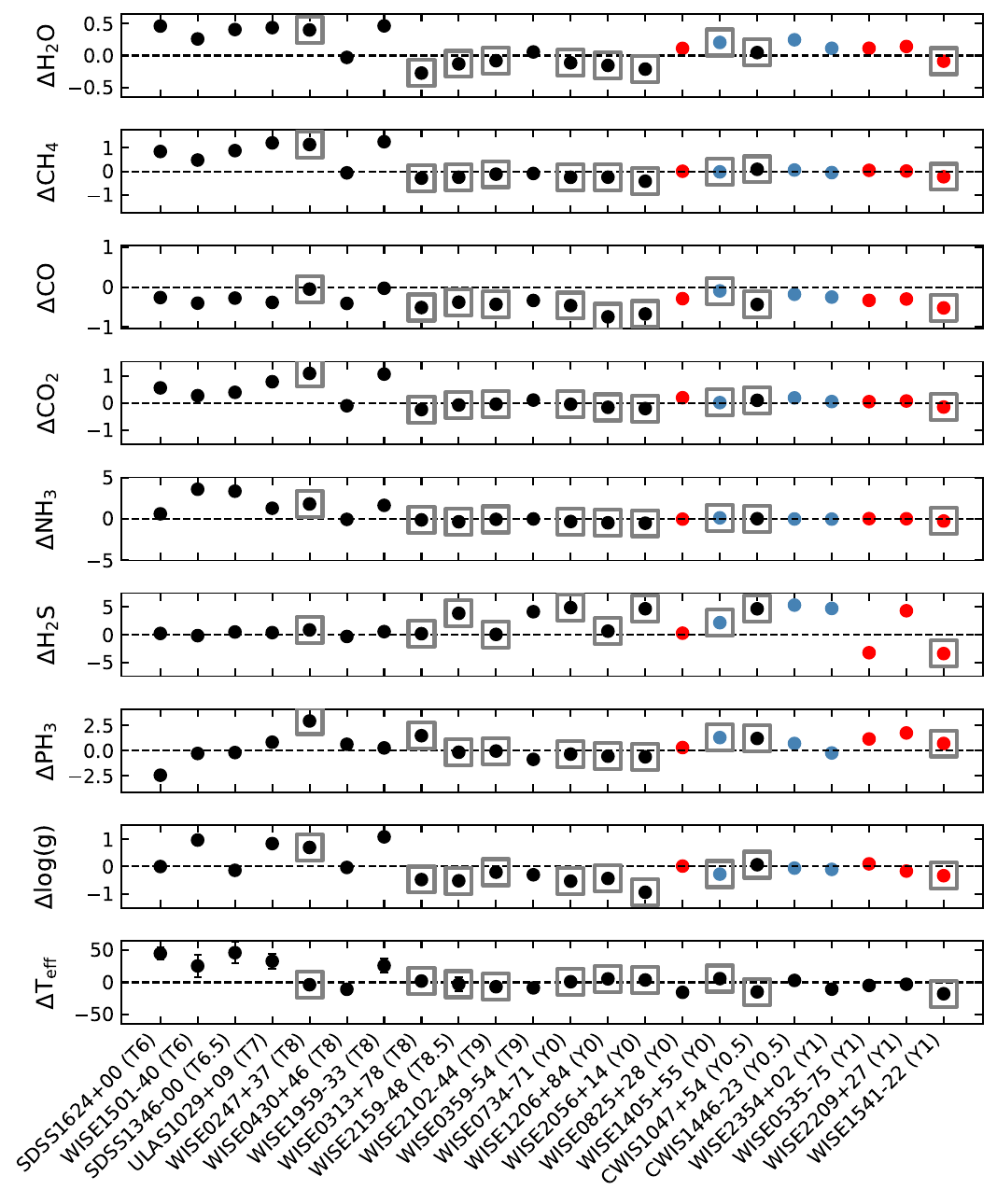}
    \caption{Each panel shows the difference between the cloudless parametric values from this work and \citet{Lueber_2026} with 1$\sigma$ uncertainties.  The data points are color-coded by model spectral fit category from this work: Regular (black), $Y$-Peculiar (blue), and $YJH$-Peculiar (red).  The grey square outlines around the data points indicate objects for which \citet{Lueber_2026} found a preference for cloudy models over cloud-free models based on Bayes factor. Note: The uncertainties for most parametric values are much smaller than their corresponding $\Delta$ and that is why that are not clearly visible.}
    \label{lueber_comparison}
\end{figure*}

To place our late-T and Y-dwarf retrieval results in the context of recent JWST-based studies, we compare our cloud-free retrievals to those reported by \citet{Lueber_2026}, who analyzed the same spectra using the Bern Atmospheric Retrieval framework (BeAR). In their analysis, \citet{Lueber_2026} compared cloud-free and gray-cloud models, quantified model preference using Bayesian evidence, and reporting Bayes factors \citep{Trotta_2008} for each object. They find that cloud-free models are generally preferred for the warmer objects (T6–T8), while cooler late-T and Y dwarfs show mixed preference, with several objects strongly favoring gray clouds (see their Table 4).

Figure~\ref{lueber_comparison} presents the object-by-object differences between our retrieved parameters and the cloud-free results from \citet{Lueber_2026}. The grey square outlines around the data points indicate objects for which \citet{Lueber_2026} find a preference for cloudy models over cloud-free models based on Bayes factors.

For \water, \methane, and \ammonia, we find that our retrieved VMRs are largely consistent with those reported by \citet{Lueber_2026} (close to $\Delta=0$), with modest object-to-object scatter. This agreement indicates that, when restricted to cloud-free models, both retrieval frameworks recover similar constraints on the primary molecular composition. The largest deviations occur among the warmer objects at the early-T end of the sample. This trend may reflect increased sensitivity to differences in thermal-profile parameterization at higher effective temperatures; however, this interpretation is speculative. A more systematic exploration of modeling assumptions and input choices will be required to determine the underlying cause of these discrepancies.

In contrast, our retrieved \co\ abundances are systematically lower than those reported by \citet{Lueber_2026} across much of the sample. This systematic trend may arise from the differences in the linelist and the thermal profile parameterization. For \coo, the two analyses are generally in agreement for the coldest objects, while discrepancies of up to $\sim$1 dex appear toward the hotter end of the sample, again suggesting sensitivity to modeling choices rather than a fundamental disagreement in the inferred atmospheric state.

The trace species H$_2$S and PH$_3$ exhibit the largest scatter in $\Delta$VMR. Several Y dwarfs show significantly higher $\Delta$H$_2$S and $\Delta$PH$_3$ values in our retrievals, while others show lower inferred abundances relative to \citet{Lueber_2026}. PH$_3$ in particular is expected to be highly sensitive to vertical mixing and disequilibrium chemistry in cold atmospheres. However, our retrieval frameworks does not explicitly incorporate vertical transport or non-equilibrium chemistry. As a result, any inferred PH$_3$ VMRs reflect an equilibrium-based forward model attempting to reproduce spectral features that may instead arise from unmodeled disequilibrium processes. Differences between studies may therefore stem from varying modeling assumptions and parameter degeneracies rather than true VMR variations. This sensitivity further underscores why we characterize our PH$_3$ constraints as tentative, particularly at the lowest effective temperatures where disequilibrium effects are expected to be most significant.

The discrepancy in surface gravity show no uniform trend between cloud-preferred and cloud-free objects; however, several of the cloud-preferred objects (square-marked points) exhibit the most negative values of $\Delta\log g$. This suggests that when clouds are favored by the data, enforcing a cloud-free model can shift opacity contributions into gravity-related parameters, although the effect is not systematic across the entire sample.

For most objects, the differences in effective temperature are small, with $\Delta T_{\mathrm{eff}}$ clustering close to zero. The largest deviations occur for the warmest objects in the sample. The close agreement in $T_{\mathrm{eff}}$ between the two studies, even for objects that prefer cloudy models in \citet{Lueber_2026}, indicates that effective temperature is robust.

Finally, the differences between the two retrieval analyses should be interpreted with caution. Residual discrepancies may arise from multiple factors, including the choice of molecular linelists, thermal profile parameterization, retrieval of gravity rather than mass, and the inclusion of additional species such as SO$_2$. Moreover, the preference for cloudy models in \citet{Lueber_2026} is based on Bayes factors, following \citet{Trotta_2008}. Recent work by \citet{thorngren_2025} has demonstrated that Bayes factors can be highly sensitive to prior choices. A more systematic exploration of prior dependence and model complexity is therefore required before drawing firm conclusions about cloud prevalence in the coldest brown dwarfs.

\section{Discussion} \label{sec:discussion}

\begin{figure*}[htb!]
    \centering
    \includegraphics[width=\linewidth]{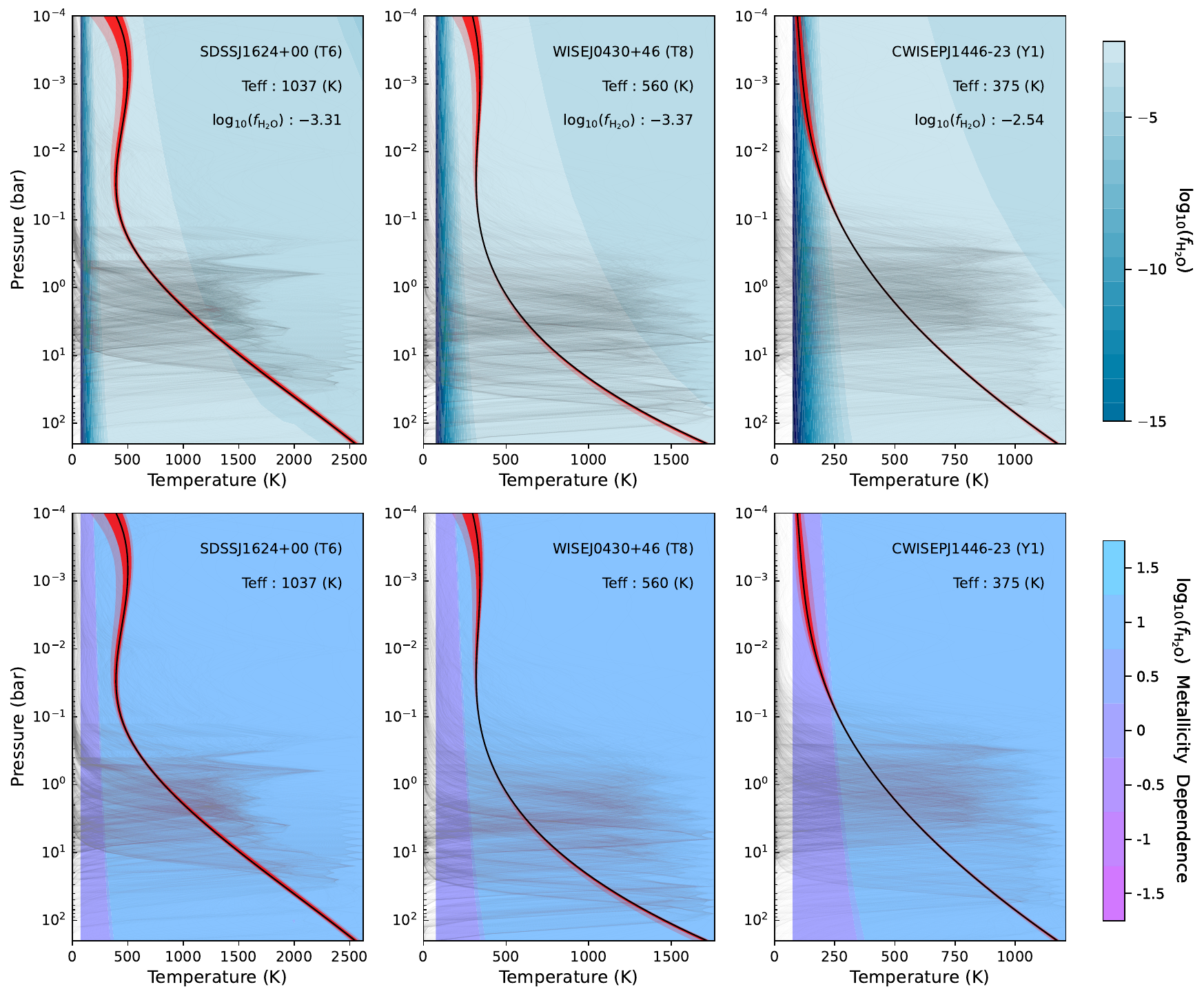}
    \caption{The panels display retrieved thermal profiles for three objects (SDSSJ1624+00 [T6], WISEJ0430+46 [T8], and CWISEPJ1446-23 [Y1]), representative of the sample, overlaid on thermochemical equilibrium \fwater\ maps at solar. The black line represents the median retrieved profile, with red shaded regions indicating the 1$\sigma$ and 2$\sigma$ central credible intervals. Lighter blue shades correspond to higher water abundances, while darker shades indicate water depletion. Grey curves are the normalized contribution functions for each object, indicating the atmospheric layers probed by the median model spectrum. Bottom: Difference between the equilibrium \fwater\ computed at [M/H]=0 and [M/H]=1 dex (C/O held fixed), with the retrieved thermal profiles overplotted. The change in \fwater\ between these two metallicities (rounded off to three significant digits) implies an approximately uniform metallicity dependence ($\sim$m$^{1}$) in the pressure–temperature region probed, so that the retrieved \fwater\ effectively provides a constraint on (O/H)$_\textrm{gas}$ under chemical equilibrium. Note: Metallicity dependence values are rounded to two significant digits.}
    \label{water_tp}
\end{figure*}

\begin{figure}[htb!]
    \centering
    \includegraphics[width=\linewidth]{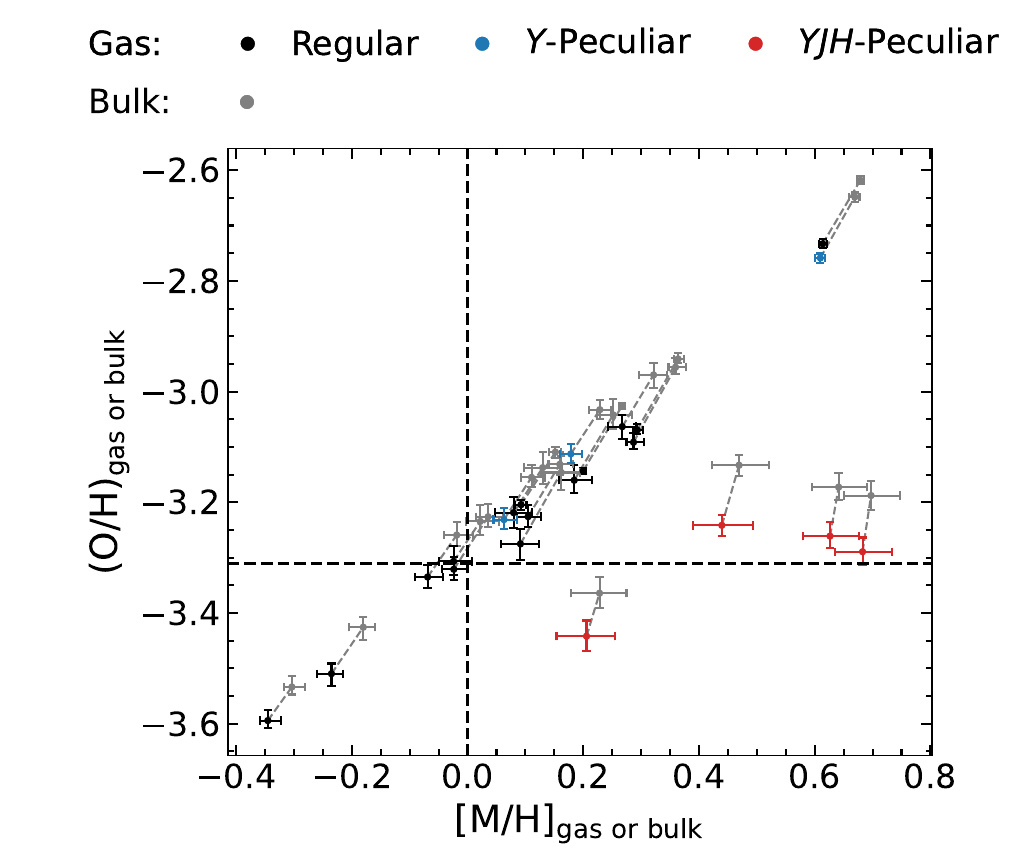}
    \caption{Calculated gas-phase (non-grey) and bulk (grey) [M/H] vs. gas-phase (non-grey) and bulk (grey) O/H ratio for our sample, with 1$\sigma$ uncertainties derived from retrieved volume mixing ratios using equations \ref{M_H}, \ref{M_H_bulk}, \ref{O_H_atm}, and \ref{O_H_bulk}, respectively. The data points are color-coded by model spectral fit category: Regular (black), $Y$-Peculiar (blue), and $YJH$-Peculiar (red). The black dashed lines represent the solar O/H ratio from \citet{Asplund_2009}.}
    \label{M_H_O_H}
\end{figure}

\begin{figure*}[htb!]
    \centering
    \includegraphics[width=\linewidth]{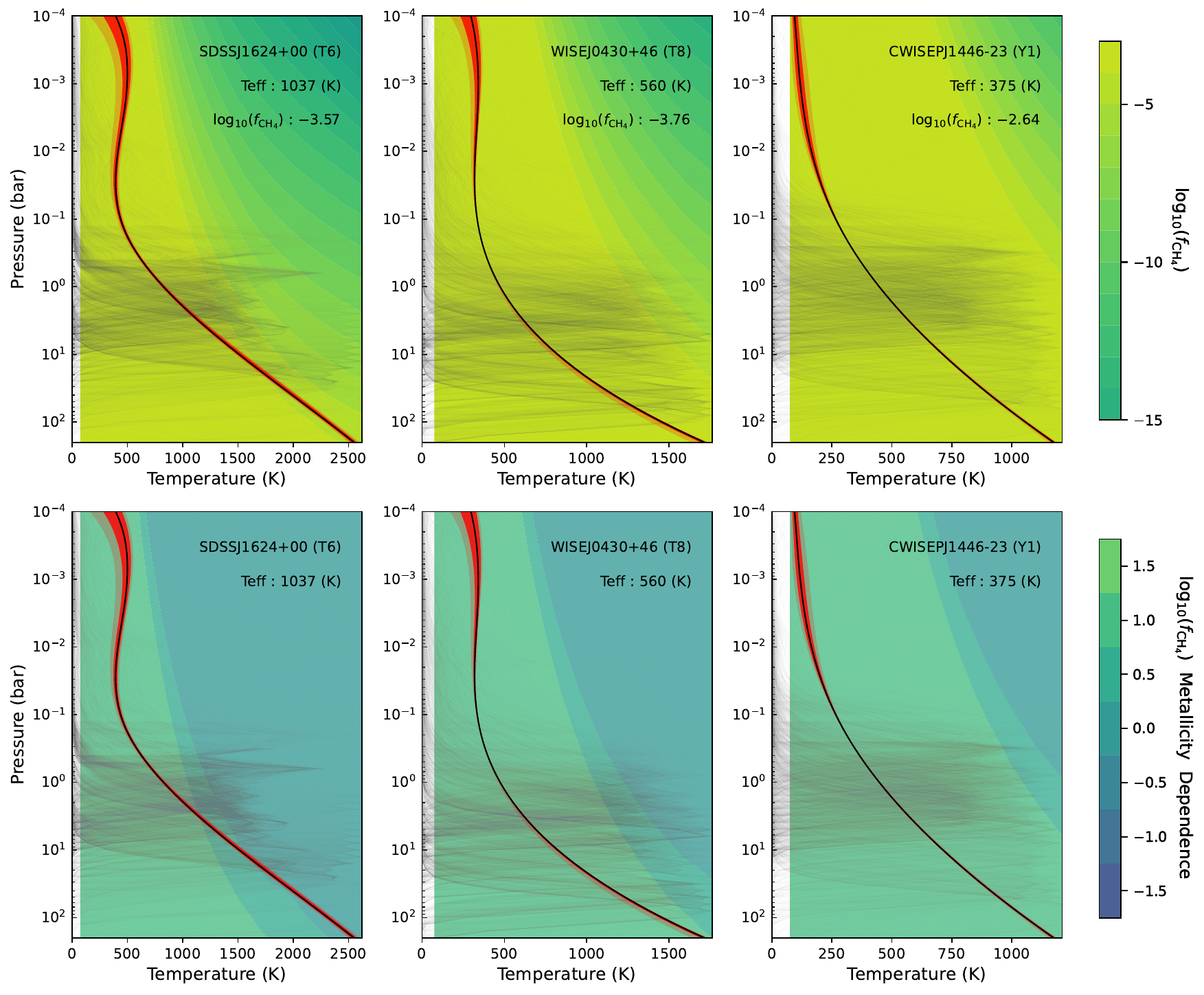}
    \caption{Top: The panels display retrieved thermal profiles for three objects (SDSSJ1624+00 [T6], WISEJ0430+46 [T8], and CWISEPJ1446-23 [Y1]), representative of the sample, overlaid on thermochemical equilibrium \fmethane\ maps at solar. The black line represents the median retrieved profile, with red shaded regions indicating the 1$\sigma$ and 2$\sigma$ central credible intervals. \fmethane\ is color-coded, with lighter yellow indicating higher VMR and darker green indicating lower VMR. Grey curves are the normalized contribution functions for each object, indicating the atmospheric layers probed by the median model spectrum. Bottom: Difference between the equilibrium \fmethane\ computed at [M/H]=0 and [M/H]=1 dex (C/O held fixed), with the retrieved thermal profiles overplotted. The change in \fmethane\ between these two metallicities (rounded off to three significant digits) implies an approximately uniform metallicity dependence ($\sim$m$^{1}$) in the pressure–temperature region probed, so that the retrieved \fmethane\  effectively provides a constraint on (C/H)$_\textrm{gas}$ under chemical equilibrium. Note: Metallicity dependence values are rounded to two significant digits.}
    \label{methane_tp}
\end{figure*}

\begin{figure}[htb!]
    \centering
    \includegraphics[width=\linewidth]{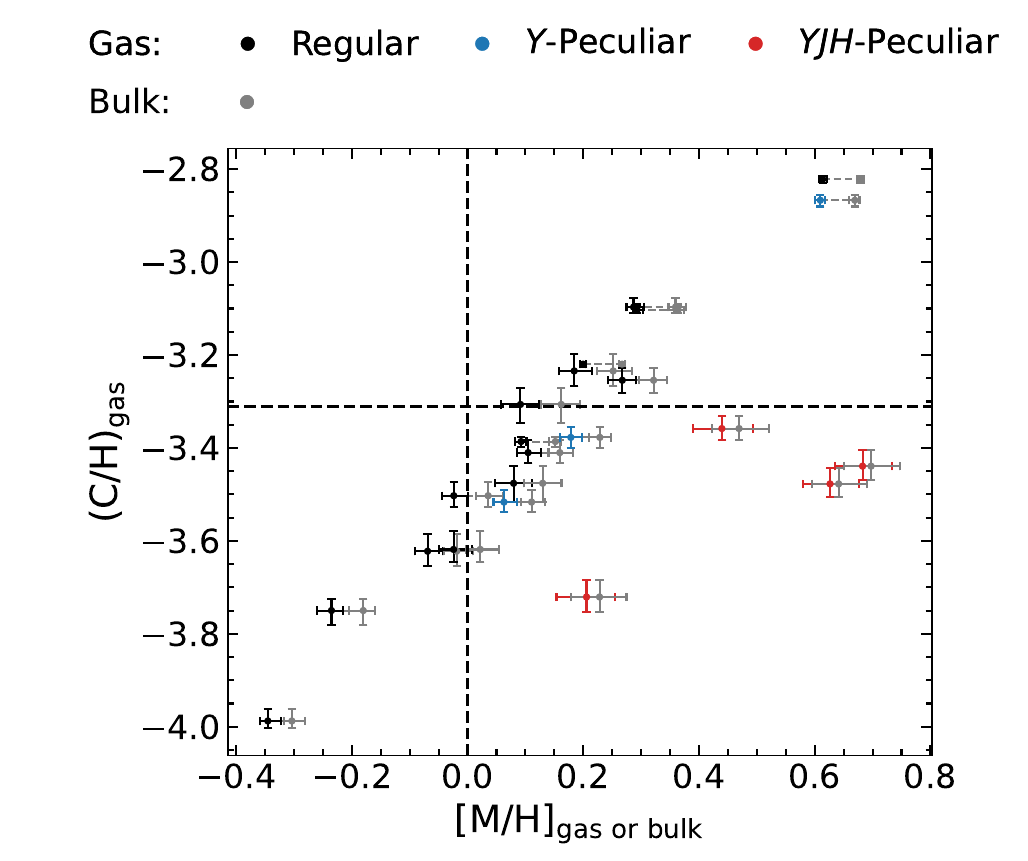}
    \caption{Calculated gas-phase (non-grey) and bulk (grey) [M/H] vs. gas-phase C/H ratio for our sample, with 1$\sigma$ uncertainties derived from retrieved volume mixing ratios using equations \ref{M_H}, \ref{M_H_bulk}, and \ref{C_H_atm}, respectively. The data points are color-coded by model spectral fit category: Regular (black), $Y$-Peculiar (blue), and $YJH$-Peculiar (red). The black dashed lines represent the solar C/H ratio from \citet{Asplund_2009}.}
    \label{M_H_C_H}
\end{figure}

We have so far focused on comparing the retrieved results from a samples perspective. In this section, we shift our focus to how the retrieved results compare with the predictions from self-consistent forward models. Specifically, we 1) compare the retrieved VMRs with the thermochemical equilibrium VMR predictions, and 2) examine the agreement between the retrieved and the forward model thermal profiles.

\subsection{Retrieved vs. Thermochemical Equilibrium VMRs}

The panels in Figure~\ref{water_tp} illustrate the retrieved thermal profiles for three representative objects spanning our \teff\ range, overlaid on thermochemical equilibrium water VMR (\fwater) maps [top] and on the corresponding metallicity–sensitivity maps [bottom].

In the top panels of Figure~\ref{water_tp}, the retrieved thermal profiles lie within iso-abundant regions of the equilibrium \fwater\ VMR ($\sim$$-4$ to $-3$), and these equilibrium VMR values closely match the retrieved \fwater\ values. The spectral sequence from T6 to Y1 shows the expected leftward shift of the thermal profiles (toward lower temperatures) with decreasing \teff, moving into regions of higher equilibrium \fwater, in agreement with equilibrium chemistry predictions \citep[e.g.,][]{Lodders_2006, Visscher_2010}, that water becomes more abundant as \teff\ decreases.

The bottom panels of Figure~\ref{water_tp} show the response of \fwater\ to metallicity, calculated by differencing the equilibrium \fwater\ computed at two metallicity grids ([M/H]$_\textrm{bulk}$ = 0 and [M/H]$_\textrm{bulk}$ = +1 dex). The resulting change in \fwater\ between the two metallicity grids is consistent with an approximately uniform scaling; in other words, increasing metallicity by +1 dex produces an approximately proportional increase in the equilibrium \fwater\ in the pressure regions probed by our spectra, as evidenced by the contribution functions.

Although subtle deviations from uniform metallicity scaling occur in the T6--T8 temperature--pressure regions, where oxygen begins to be sequestered into condensates such as silicates and sulfides, these effects are modest along the portions of the profiles that contribute most strongly to the emergent flux. In the plateau regions, oxygen remains predominantly in gaseous \water. Under the assumptions of chemical equilibrium and fixed C/O, the retrieved \fwater\ in these plateau regions therefore serve as an effective tracer of (O/H)$_\textrm{gas or bulk}$. As shown in Figure~\ref{M_H_O_H}, (O/H)$_\textrm{gas or bulk}$ increases with [M/H]$_\textrm{gas or bulk}$, revealing a positive trend in which atmospheres with greater bulk metal enrichment also contain enhanced oxygen abundances. This trend is expected as \fwater\ is the dominant oxygen-bearing chemical species and responds proportionally to metallicity, especially in the pressure regions sampled by the data (see Figure \ref{H2O_CH4_M_H} and \ref{water_tp}). Most objects in our sample are therefore inferred to be enriched in both oxygen and overall metals relative to solar composition.

Figure~\ref{methane_tp} illustrates an analogous analysis for methane. The top panels show the retrieved thermal profiles plotted on equilibrium methane VMR (\fmethane) maps. As \teff\ decreases along the T/Y sequence, the thermal profiles again shift leftward toward lower temperatures, moving into regions of higher equilibrium \fmethane. This trend aligns with thermochemical equilibrium predictions, in which carbon chemistry increasingly favors \methane\ at cooler temperatures. In T dwarfs (\teff\ $\gtrsim$ 800 K), \methane\ is less abundant at high pressures ($\gtrsim$10 bar) than at low pressures ($\lesssim$1 bar), consistent with carbon being preferentially locked in \co\ at higher temperatures and deeper layers. Across the profiles, the retrieved temperatures tend to occupy regions where \fmethane\ is lower than \fwater, again in line with the expectation that \water\ is the most abundant molecular species, followed by \methane, in these cold atmospheres.

The bottom panels of Figure~\ref{methane_tp} show the metallicity dependence of \fmethane, again calculated by differencing thermochemical equilibrium \fmethane\ at [M/H] = 0 and [M/H] = +1 dex. Like with \fwater, the change in \fmethane\ between the two metallicity grids are broadly consistent with an mostly uniform scaling (power-law exponent $\sim$1) over the pressures probed by our spectra. In the regions where the contribution functions peak, the metallicity dependence maps again exhibit a relatively uniform scaling with temperature and pressure, indicating that \fmethane\ responds nearly proportionally to changes in overall metal content.

However, for \methane\ there is a crossover in metallicity dependence between the T6 and T8 thermal profiles. This arises because the thermal profiles for the objects lie on opposite sides of the CO–\methane\ transition. For the warmer T6 dwarf, a fraction of carbon remains partitioned in CO at the relevant temperatures and pressures, partially moderating the response of \fmethane\ to changes in metallicity. In contrast, the cooler T8 dwarf has effectively transitioned to the \methane-dominated regime, where nearly all carbon resides in \methane; in this limit, the methane abundance scales more uniformly with metallicity and therefore serves as an effective tracer of (C/H)$_\textrm{gas}$. Figure~\ref{M_H_C_H} shows a positive correlation between (C/H)$_\textrm{gas}$ and [M/H]$_\textrm{gas or bulk}$, indicating that atmospheres with higher overall metal enrichment also exhibit enhanced carbon abundances.  This trend is expected as \fmethane\ is the dominant carbon-bearing chemical species and responds proportionally to metallicity, especially in the pressure regions sampled by the data (see Figure \ref{H2O_CH4_M_H} and \ref{methane_tp}). Most objects in our sample therefore appear enriched in both carbon and overall metals relative to the solar composition.

Taken together, these results justify the interpretation adopted in \S\ref{metallicity}: in the temperature–pressure regions sampled by our spectra, the retrieved \fwater\ and \fmethane\ primarily reflect the underlying elemental enrichment [(O/H) and (C/H)$_\textrm{gas}$, respectively] rather than local thermal variations or condensation processes. The iso-abundance plateaus in Figures~\ref{water_tp} and \ref{methane_tp} [top], combined with the uniform metallicity dependence in Figures~\ref{water_tp} and \ref{methane_tp} [bottom] demonstrate that both molecules behave as effective tracers of bulk metal content across the T/Y sequence under the assumptions of chemical equilibrium and fixed C/O.

\subsection{Retrieved vs. Forward Model Thermal Profile}

The panels in Figures \ref{tp1}–\ref{tp4} compare the retrieved thermal profiles with those from the cloudless Elf-Owl forward model \citep{Mukherjee_2024_sonora} for all objects in our sample. For objects with \teff\ $>$ 900 K, the retrieved thermal profiles are hotter than the Elf-Owl thermal profiles at greater than 0.1 bar, which is the region that our observed spectra probes. Objects with \teff\ less than 750 K, except for WISEPA J0313+78, have retrieved thermal profiles that are colder than Elf-Owl thermal profiles at pressures greater than 1 bar. \citet{Kothari_2024} demonstrated that a five-knot spline parameterization can reproduce the Elf-Owl thermal profile well, with a root-mean-square deviation of 17 K for a Y dwarf. Therefore, the differences we are seeing between the retrieved and the forward model thermal profile are real and statistically significant.

To further evaluate the performance of our retrieval framework, we ran a retrieval on a forward model spectrum. {We adopted a similar Elf-Owl forward model spectrum (\teff\ of 450 K, \logg\ of 3.23 cm/s$^2$, C/O of 0.5, and [M/H] of 0.0) to WISE 0359$-$54 from \citet{Beiler_2023}. We then convolved it to the resolution of the observed spectrum, and assigned it the same noise properties as the data. Using this synthetic spectrum as input, we then perform a retrieval following the setup described in \S3.

We find that the Elf-Owl retrieved spectrum does reproduce the input Elf-Owl forward model very well (see Figure \ref{spectrum_EO}.). Despite this agreement between the retrieved and the Elf-Owl spectrum, the retrieved thermal profile differ from the Elf-Owl profile as shown in Figure~\ref{EO_TP_compare}. The retrieved thermal profile is hotter by up to $\sim$100 K between 0.1 to 1 bar in pressure, which is also the region that is most probed by the spectrum as shown by the grey contribution functions. This discrepancy is caused by the difference in the surface gravity from the Elf-Owl forward model, which has a \logg\ 3.23 [cm/s$^2$],  whereas the retrieval favors a higher surface gravity of $\sim$4 [cm s$^{-2}$], driven by its preference for a higher mass. As for why the retrievals prefer a higher mass and by extension a higher surface gravity compared to the forward models is an active area of research.

The key difference between the retrieval and the forward model arise from the treatment of chemistry. The Elf-Owl models adopt rainout equilibrium chemistry, in which condensates such as silicates and alkalis are removed from the gas-phase as they condense, resulting in depleted gas-phase Na and K abundances at lower temperatures and pressures. In contrast, our retrieval framework assumes vertically uniform abundances for all species, without enforcing chemical equilibrium or rainout. As a result, the retrieval is free to adjust alkali abundances to fit the data and potentially compensate for missing condensation effects by modifying the local thermal gradient, which can propagate into differences in the inferred surface gravity. Therefore, part of the thermal profile discrepancy likely reflects the contrasting chemistry assumptions and abundance parameterization. Future work with pressure-dependent alkali abundances or condensation-informed parameterizations (e.g., rainout-chemistry informed priors, etc.) will need to be tested to see if they can better capture condensation effects and explain the discrepancy between the forward model and retrieval results.

\section{Acknowledgement}

This work is based [in part] on observations made with the NASA/ESA/CSA James Webb Space Telescope. The data were obtained from the Mikulski Archive for Space Telescopes at the Space Telescope Science Institute, which is operated by the Association of Universities for Research in Astronomy, Inc., under NASA contract NAS 5-03127 for JWST. The specific observations analyzed can be accessed via doi:10.17909/ntwg-k441n and are associated with program \#2302. This research is based on observations made with the NASA/ESA Hubble Space Telescope obtained from the Space Telescope Science Institute, which is operated by the Association of Universities for Research in Astronomy, Inc., under NASA contract NAS 5–26555. These observations are associated with program(s) 12330, 12544, 12970, and 13178. The calculations required for the results of this paper were done using the Owens cluster \citep{Owens_2016} at the Ohio Supercomputer Center \citep{OSC_1987}. Ben Burningham acknowledges support from UK Research and Innovation Science and Technology Facilities Council [ST/X001091/1]. The research shown here acknowledges use of the Hypatia Catalog, an online compilation of stellar abundance data as described in \citep{Hinkel_2014}. We would also like to thank the anonymous referee for a careful reading of the manuscript and for thoughtful comments that helped improve the clarity and quality of this work.

\software{Corner \citep{corner_github},
Matplotlib \citep{matplotlib},
Numpy \citep{NumPy-Array}}

\section{Appendix}

We calculate the gas-phase oxygen-to-hydrogen ratio (O/H)$_\textrm{gas}$  using:

\begin{equation}
(\textrm{O}/\textrm{H})_\textrm{gas} =
\frac{f_{\mathrm{H_2O}} + 2f_{\textrm{CO}_{2}} + f_{\textrm{CO}}}{%
2\left([1-\sum f_\textrm{metals}] \times 0.84\right)+\sum n f_{{\rm H}_n}},
\label{O_H_atm}
\end{equation}

\noindent where the numerator accounts the VMR for all the oxygen-bearing species. The denominator represents the total atmospheric hydrogen content, as described in \S\ref{metallicity}. The bulk oxygen-to-hydrogen ratio is calculated using:

\begin{equation}
(\textrm{O}/\textrm{H})_\textrm{bulk} =
\frac{\mathrm{O}_\mathrm{bulk}}{%
2\left([1-\sum f_\textrm{metals}] \times 0.84\right)+\sum n f_{{\rm H}_n}},
\label{O_H_bulk}
\end{equation}

\noindent where $\mathrm{O}_\mathrm{bulk}$ is $\mathrm{O_{\mathrm{gas}}}\times0.371\times(\mathrm{C}/\mathrm{O})_\mathrm{bulk}$ (see \S\ref{metallicity} for a more detailed calculation).

We calculate the gas-phase carbon-to-hydrogen ratio (C/H)$_\textrm{gas}$ using:

\begin{equation}
(\textrm{C}/\textrm{H})_\textrm{gas} =
\frac{f_{\textrm{CO}} + f_{\textrm{CO}_{2}} + f_{\textrm{CH}_{4}}}{%
2\left([1-\sum f_\textrm{metals}] \times 0.84\right)+\sum n f_{{\rm H}_n}}
\label{C_H_atm}
\end{equation}

\noindent where the numerator accounts the VMR for all the carbon-bearing species. The denominator represents the total atmospheric hydrogen content, as described in \S\ref{metallicity}.

\begin{figure*}[htb!]
    \centering
    \includegraphics[width=\linewidth]{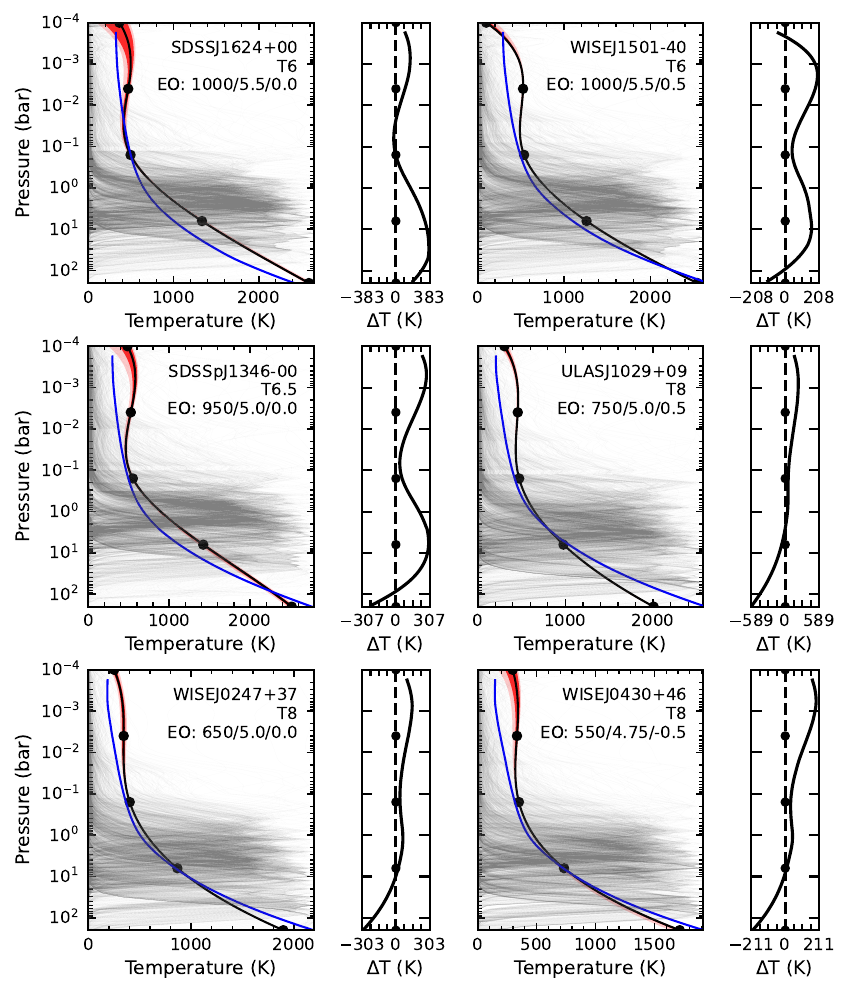}
    \caption{Retrieved and forward-model thermal profiles comparison for six brown dwarfs. In each row, the first and third panel show the retrieved and forward model thermal profile. The black line is the median retrieved thermal profile and the red regions around it represent the 1$\sigma$ \& 2$\sigma$ central credible interval. The blue line represents the Elf-Owl thermal profile that closely matches the \teff (K), \logg [cms$^{-2}$], [M/H]$_\textrm{gas}$, and (C/O) (0.5) of the object. The second and fourth panels in each row shows the difference between the retrieved and Elf-Owl thermal profile. The black dots in each panel represent the position of the five knots.}
    \label{tp1}
\end{figure*}

\begin{figure*}[htb!]
    \centering
    \includegraphics[width=\linewidth]{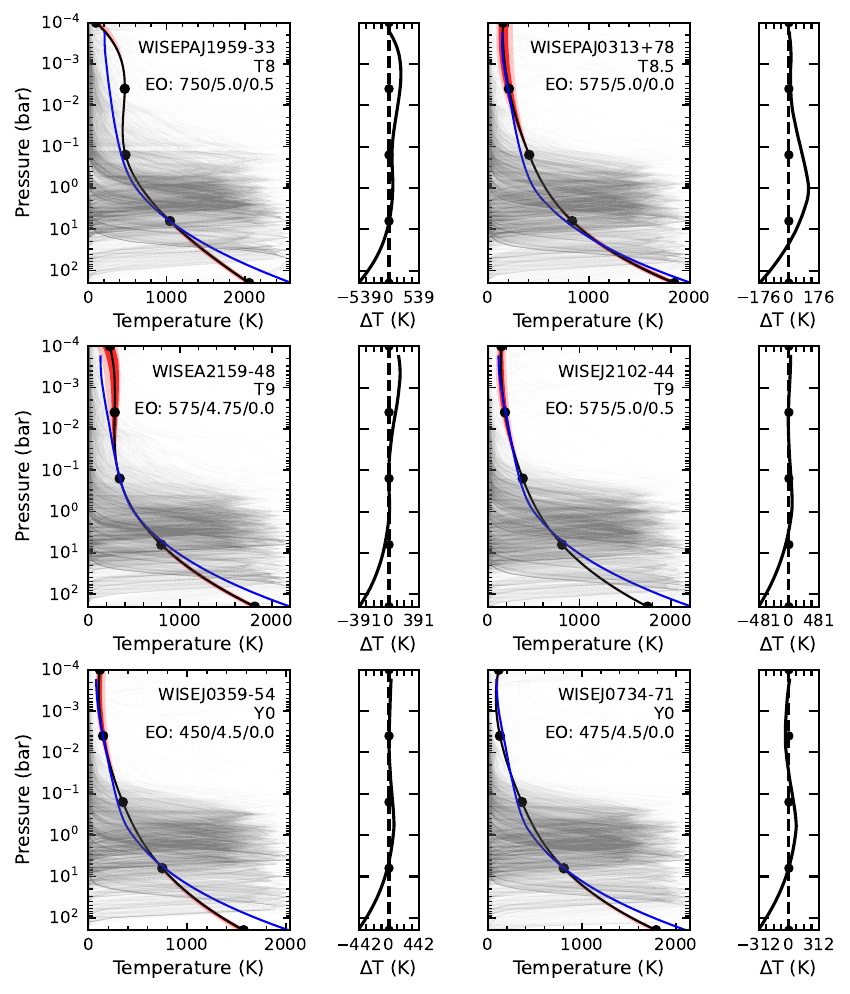}
    \caption{Retrieved and forward-model thermal profiles comparison for six brown dwarfs. In each row, the first and third panel show the retrieved and forward model thermal profile. The black line is the median retrieved thermal profile and the red regions around it represent the 1$\sigma$ \& 2$\sigma$ central credible interval. The blue line represents the Elf-Owl thermal profile that closely matches the \teff (K), \logg [cms$^{-2}$], [M/H]$_\textrm{gas}$, and (C/O) (0.5) of the object. The second and fourth panels in each row shows the difference between the retrieved and Elf-Owl thermal profile. The black dots in each panel represent the position of the five knots.}
    \label{tp2}
\end{figure*}

\begin{figure*}[htb!]
    \centering
    \includegraphics[width=\linewidth]{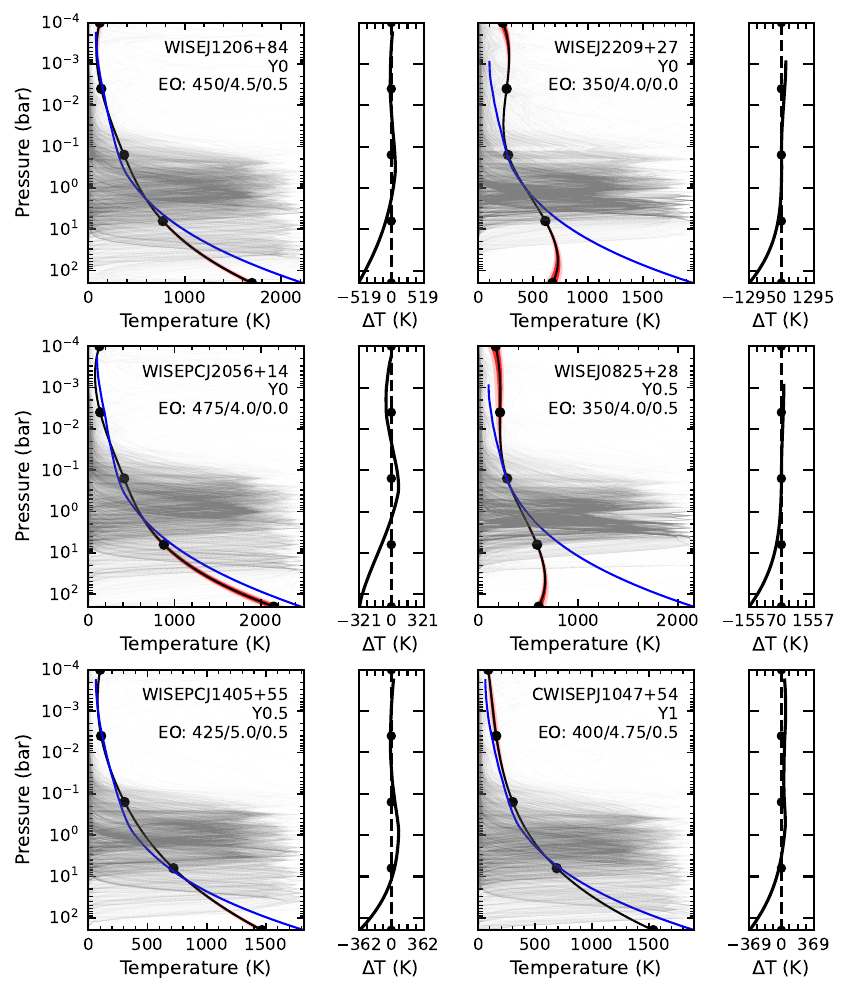}
    \caption{Retrieved and forward-model thermal profiles comparison for six brown dwarfs. In each row, the first and third panel show the retrieved and forward model thermal profile. The black line is the median retrieved thermal profile and the red regions around it represent the 1$\sigma$ \& 2$\sigma$ central credible interval. The blue line represents the Elf-Owl thermal profile that closely matches the \teff (K), \logg [cms$^{-2}$], [M/H]$_\textrm{gas}$, and (C/O) (0.5) of the object. The second and fourth panels in each row shows the difference between the retrieved and Elf-Owl thermal profile. The black dots in each panel represent the position of the five knots.}
    \label{tp3}
\end{figure*}

\begin{figure*}[htb!]
    \centering
    \includegraphics[width=\linewidth]{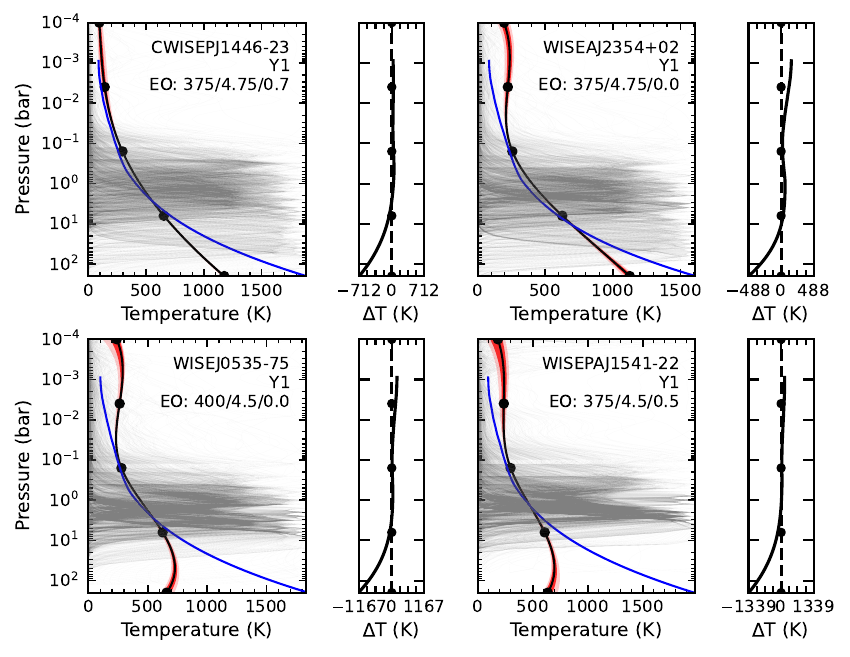}
    \caption{Retrieved and forward-model thermal profiles comparison for four brown dwarfs. In each row, the first and third panel show the retrieved and forward model thermal profile. The black line is the median retrieved thermal profile and the red regions around it represent the 1$\sigma$ \& 2$\sigma$ central credible interval. The blue line represents the Elf-Owl thermal profile that closely matches the \teff (K), \logg [cms$^{-2}$], [M/H]$_\textrm{gas}$, and (C/O) (0.5) of the object. The second and fourth panels in each row shows the difference between the retrieved and Elf-Owl thermal profile. The black dots in each panel represent the position of the five knots.}
    \label{tp4}
\end{figure*}

\begin{figure*}[htb!]
    \centering
    \includegraphics[width=\linewidth]{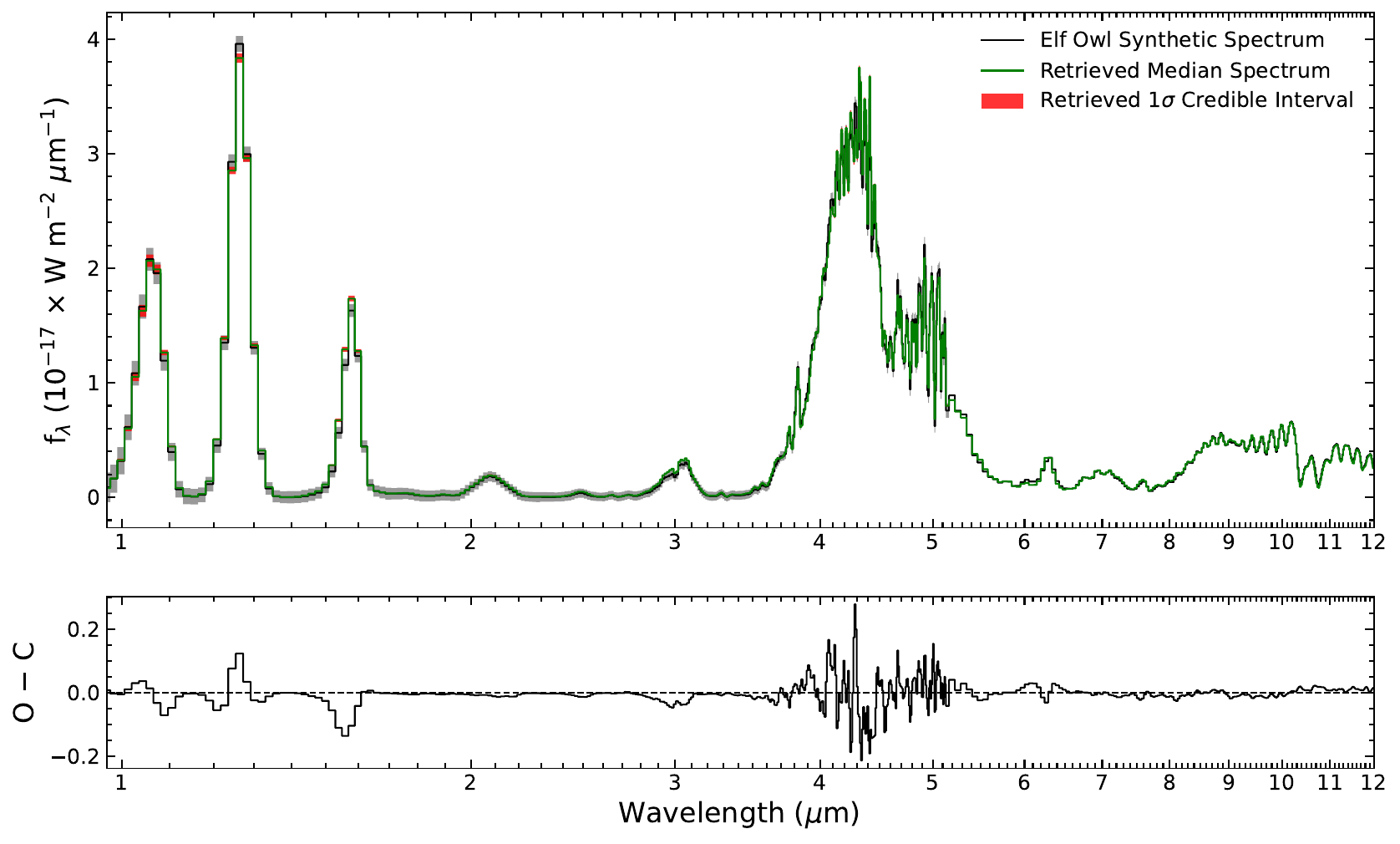}
    \caption{The top panel shows the Elf-Owl forward model spectrum at a \teff\ of 450 (K), \logg\ of 3.23 [cm/s$^2$], C/O of 0.5, and [M/H] of 0.0 in black covering $\sim$0.97–12.1 $\mu$m with 1$\sigma$ uncertainties in grey in units of f$_{\lambda}$ adapted from the observed spectrum uncertainties of WISE 0359--54. The retrieved median spectrum is shown in green and the red region shows the 1$\sigma$ central credible interval around the median spectrum. The bottom panel shows the difference between the observed spectrum (O) and the retrieved median spectrum (C).}
    \label{spectrum_EO}
\end{figure*}

\begin{figure}[htb!]
    \centering
    \includegraphics[width=\linewidth]{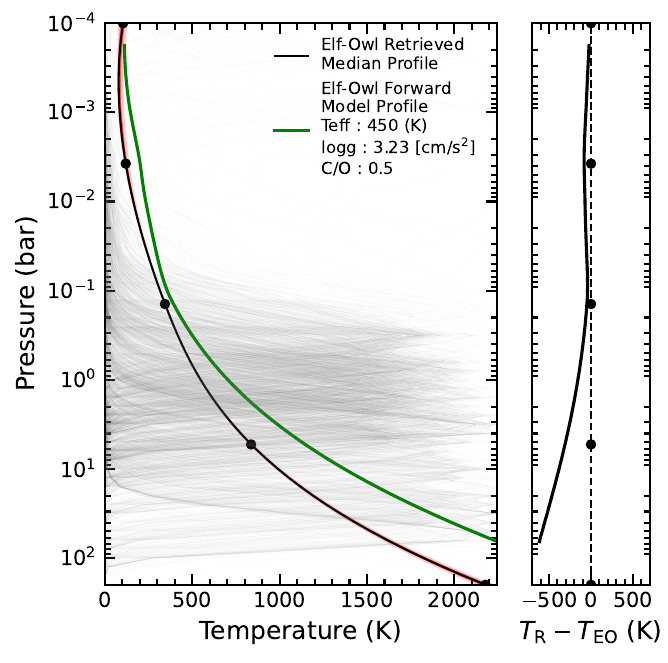}
    \caption{Left Panel: Retrieved versus forward-model Elf-Owl thermal profile. The black line shows the retrieved median thermal profile, with dark and light red regions indicating the 1$\sigma$ and 2$\sigma$ central credible intervals, respectively. The grey lines represent normalized contribution functions, illustrating the atmospheric layers probed by the spectrum, and the black dots represent the five retrieved knots. The green line represents the Elf-Owl thermal profile at a \teff\ of 450 (K), \logg\ of 3.23 [cm/s$^2$], C/O of 0.5, and [M/H] of 0.0. Right Panel: Shows the difference between the retrieved median and the Elf-Owl forward model thermal profile}
    \label{EO_TP_compare}
\end{figure}

\newpage

\bibliographystyle{aasjournal}
\bibliography{Reference}

\end{document}